\documentclass[fleqn,usenatbib]{mnras}

\usepackage{newtxtext,newtxmath}
\usepackage[normalem]{ulem}

\usepackage{color}
\usepackage{microtype}

\usepackage[OT2,T1]{fontenc}
\usepackage{ae,aecompl}
\usepackage{graphicx}	
\usepackage{amsmath}	
\usepackage{amssymb,bm}	

\DeclareSymbolFont{cyrletters}{OT2}{wncyr}{m}{n}
\DeclareMathSymbol{\Sha}{\mathalpha}{cyrletters}{"58}

\title[VLA L-band beam modelling]{Primary beam effects of
  radio astronomy antennas: I. Modelling the Karl G. Jansky Very Large
  Array (VLA) L-band beam using holography}

   \author[K. Iheanetu et al.]{K. Iheanetu$^{1}$\thanks{E-mail: g15i9152@campus.ru.ac.za},
J. N. Girard$^{2,1}$, 
          O. Smirnov$^{1,3}$,
          K. M. B. Asad$^{3,1,4}$,
          M. de Villiers$^{3}$,
          \newauthor
          K. Thorat$^{1,3}$,  
          S. Makhathini$^{3,1}$,
          R. A. Perley$^{5,1}$
         \\
              $^{1}$Department of Physics and Electronics, Rhodes University, PO Box 94, Grahamstown, 6140, South Africa\\
              $^{2}$AIM, CEA, CNRS, Université Paris-Saclay, Université Paris Diderot, Sorbonne Paris Cité, F-91191 Gif-sur-Yvette, France\\
              $^{3}$South African Radio Astronomy Observatory, 3rd Floor, The Park, Park Road, Pinelands, Cape Town, 7405, South Africa\\
              $^{4}$Department of Physics and Astronomy, University of the Western Cape, Bellville, Cape Town, 7535, South Africa\\
             $^{5}$NRAO, Charlottesville, USA
        }
   \date{Received xxxx; accepted yyy}
\pubyear{2018}

\begin{document} 
\maketitle


\begin{abstract}
Modern interferometric imaging relies on advanced calibration that
incorporates direction-dependent effects.  Their increasing number of
antennas (e.g. in LOFAR, VLA, MeerKAT/SKA) and sensitivity are often
tempered with the accuracy of their calibration. Beam accuracy drives
particularly the capability for high dynamic range imaging (HDR --
contrast > 1:10$^6$). The Radio Interferometric Measurement Equation
(RIME) proposes a refined calibration framework for wide field of
views (i.e. beyond the primary lobe and first null) using beam
models. We have used holography data taken on 12 antennas of the Very
Large Array (VLA) with two different approaches: a `data-driven'
representation derived from Principal Component Analysis (PCA) and a
projection on the Zernike polynomials. We determined sparse
representations of the beam to encode its spatial and spectral
variations. For each approach, we compressed the spatial and spectral
distribution of coefficients using low-rank approximations. The
spectral behaviour was encoded with a Discrete Cosine Transform
(DCT). We compared our modelling to that of the \emph{Cassbeam}
software which provides a parametric model of the antenna and its
radiated field. We present comparisons of the beam reconstruction
fidelity vs. `compressibility'. We found that the PCA method provides
the most accurate model. In the case of VLA antennas, we discuss the
frequency ripple over L-band which is associated with a
standing wave between antenna reflectors. The results are a series of
coefficients that can easily be used `on-the-fly' in calibration
pipelines to generate accurate beams at low computing costs.

\end{abstract}
\begin{keywords}
Instrumentation: interferometers - methods: analytical - methods: data analysis
\end{keywords}



\section{Introduction}
\label{sec-intro}
New large radio interferometers, such as the Low-Frequency Array
(LOFAR -- \citep{LOFAR}) and the Square Kilometre Array (SKA --
\citep{SKA}), bring the potential of very high angular, temporal and
spectral resolutions, along with improved raw sensitivity. However,
faster, more efficient and more accurate methods are required to
tackle the data storage and calibration challenges imposed by the size
and complexity of these instruments. In addition, the quality of the
scientific products strongly depends on one's ability to cope with the
instrumental and environmental distortions occurring during an
observation. In order to detect faint sources, measure accurate flux
density in catalogues from survey data, study extended emission, such
effects are quickly becoming the limiting factors that limit the
actual performance of such instruments.

\subsection{New calibration framework to improve imaging}
Due to the very large field of view (FOV) of the aforementioned
instruments, the `classical' calibration and imaging methods reach
their limits in quality. \cite{hamaker96} proposed a new framework
that efficiently formulates the radio-polarimetric inverse
problem. This mathematical framework, known as the Radio
Interferometer Measurement Equation (RIME, see review by
\cite{smirnov_2011} and references therein), enables a way to
formulate the corruptions of the electric field coming from an
astronomical source by a series of linear effects (the `Jones'
matrices) taking place between the source and the observer. Such a
framework facilitates improved modelling and calibration of complex
direction-independent effects (DIE, such as antenna gain, Faraday
rotation, etc.) and direction-dependent effects (DDE, such as antenna
primary beam (PB) and ionospheric effects).  In order to achieve high
dynamic range (HDR) in radio interferometric images (reaching nowadays
contrast ratios $>10^6$:1), knowledge of the antenna response in the
array is critical. An incorrect beam model will decrease the accuracy
of images of off-axis sources that are far from the phase centre and
will lead to strong distortions that further time or frequency
integration cannot eliminate.  Indeed, the primary beam induces
temporal variations (due to parallactic angle rotation) and frequency
variations of the source brightness. As a result, towards the edge of
the antenna FOV and in the secondary lobes of the antenna pattern,
radio sources with constant flux density can appear as falsely varying
sources which interfere with the calibration process that assumes a
steady sky model. The standard calibration procedure tries to
compensate for these apparent variable sources by generating incorrect
calibration solutions, which then lead to strong residual artefacts
around strong sources, preventing the detection of faint sources
hidden in the calibration noise.

\subsection{Why beam models are vital?}
The purpose of this study is to propose methods to model realistic
antenna beams to be used in the calibration process for solving for
DDEs. Knowing the realistic behaviour of each antenna beam pattern,
including beyond the first null, is vital for reaching HDR with
minimal bias. Direction-dependent calibration techniques (like `source
peeling', differential gains calibration, etc.) can correct for
artefacts, but the resulting flux suppression compromises the data
fidelity.  The goals for this study are:
\begin{itemize} 
\item to obtain an accurate beam representation that suits HDR imaging.
\item to produce a model of the beam beyond its first null (a work
  which is currently not done extensively in the community)
\item to develop a telescope-independent set of methods that capture
  the spatial and spectral morphology of the beam in the most generic
  way.
\item to provide a model with the smallest number of degrees of
  freedom, in order to only capture the relevant components of the
  beam and its variations.
\end{itemize}
The power of the RIME framework enables solution for arbitrary
specific DDE parameters. Once an accurate representation of the beam
is obtained, it can then be turned into a parametric model (Jones)
matrix of known structure that can be solved for during the
calibration process. There is scope for precise beam modelling that
can contribute to the calibration procedure by giving a very accurate
first estimation of the realistic beam. The interest of defining such
models is double: i) it enhances current DDE calibration with
`refined' a priori knowledge of the beam model, and ii) it enhances
the search for a parametric and generic model (for this antenna
geometry) that can be directly solved for during DDE calibration. It
will not only improve the quality of images but also give useful
feedback on the state of the instrument. Studies in this domain are
relatively not common in the community and our work aims at
contributing to it.

In this paper (the first in a series), we propose two complementary
approaches to address beam modelling using sparse representations
derived from holographic measurements taken at the VLA in 2016. It is
organized as follows: the main principle behind the beam holography
measurements taken at the VLA in introduced in Sec. \ref{sec-holo},
Sec. \ref{sec-beamanalysis} describes the main characteristics of the
measured beam, compared to a theoretical electromagnetic (EM)
model. Sec. \ref{sec-beammodelling} describes the modelling of this
observed beam using two approaches, the first one with Principal
Component Analysis (PCA) and Singular Value Decomposition (SVD), and
the second one with the Zernike polynomials that suit the shape of the
antenna electric field distribution.

\section{Radio holographic measurements}
\label{sec-holo}
\subsection{Principles of radio holography}
The aim of radio holographic measurements with an interferometer is to
measure the normalized complex field of the primary beam (or simply
beam) of a radio antenna. In this context, `Normalized' means that the resulting
measurements are relative to the antenna beam's on-axis values, which
are nominally given an amplitude of 1.0, and a phase of 0.  Radio
holography is done by correlating the voltages from `fixed' antennas
with those from `moving' antennas.  The `fixed', or `reference',
antennas track a strong point source, while the `moving' antennas
sample the antenna beam by stepping through an angular ($l$, $m$) grid
centered on the (moving) radio source.  The complex product between a
`moving' and `fixed' antenna, following calibration, directly provides
measures of the normalized voltage beam.

It is shown that Fourier transformation of the complex beam results in an image of the effective antenna aperture, which can then be analyzed for deviations in an antenna's panel settings \citep{baars2007near}. This is the
normal application of the radio holography methodology, and has been
used by many instruments for improving the efficiency of the antennas.
However, for this work, we utilize the antenna beam measurements
directly, as is is these that we wish to parameterize.

\subsection{VLA holography measurements}
\label{sec-holovla}
The holography methodology has been developed by VLA staff over the
past 20 years in order to enable panel adjustments for improved high
frequency performance.    The holography program has resulted in
spectacular improvements in high-frequency VLA performance.  In normal
holography, (where the goal is to measure the reflector
perturbations) the beam is scanned in an N x N grid, with stepsize
set to 1/1.2 times the critical sampling value of $\Delta l =
\lambda/D$ radians (i.e., a 20\% oversampling).  The maximum offset
angle, $L = N\Delta l/2.4$ is related to the physical resolution of
the transform by roughly $\delta x \sim 1.27\lambda/L$.

However, the near-critical sampling of the antenna beam which is
sufficient for panel holography is not well suited to generating
accurate measures of the antenna beam.  For this purpose (`beam
holography') the holography mode of the VLA was modified to enable measurement of the primary beam (e.g. \cite{Cotton_1994}). To achieve this, it is necessary to much more densely sample the antenna beam.  Experience indicates that an oversampling factor of about 5 is
necessary to permit accurate interpolation of the antenna beam.  


The observations utilized in this work met the following criteria:
\begin{enumerate}
\item Source selection and timing: The strong radio source 3C147
  (J0542+4951) was chosen.  Its flux density of $\sim$22 Jy at 1.4 GHz
  is sufficient to accurately measure the beam out to the
  second null with adequate SNR.  The square-grid sampling scheme
  necessary for holography causes excessive antenna motion at low
  frequencyies near meridian transit for sources like 3C147 which
  transit near the zenith.  Thus, these observations were taken after
  transit, starting at an elevation of 60 degrees, and ending at an
  elevation of 22 degrees.  
\item Raster step size, extent, and dwell time: A step size of 309
  arcseconds was chosen to adequately oversample the beam at the
  high-frequency end of the band (2.0 GHz).  At this frequency, the
  oversampling factor is 4.0. At the low-frequency end (1.0 GHz), the
  oversampling is now 8.0, so a raster size of 35 x 35 was chosen in
  order to ensure coverage of the primary beam to the second null. The
  primary raster motion was in the vertical -- starting at one corner
  of the grid, the moving antennas scanned downwards at constant $l$
  offset, then upwards at the next $l$ offset.  Each vertical scan is
  360 seconds long (35 points x 10 seconds, plus 10 seconds settling
  time at the beginning).  A dwell time, per raster point, of 10 seconds
  was chosen to give sufficient SNR of the inner sidelobes. Due to the
  antenna motion between steps, six seconds of actual integration
  time, per raster point, was utilized. The total time taken for these
  observations was 4.1 hours.
\item Array configuration and reference antenna selection: To ensure
  minimal cross-coupling between antennas (which is an important
  problem at low frequencies with short spacings), the observations
  were taken in the A-configuration.  Since excellent phase stability
  is required in holography, the `reference' and `target' antennas
  alternated down each arm of the array, ensuring that any given
  `target' antenna had two adjacent reference antennas, one on each
  side.  Thus, twelve of the antennas in the array were scanned -- 5,
  6, 8, 10, 12, 14, 15, 20, 23, 26, 27, 28.  The remaining antennas
  were either reference, or were out of the array at the time of
  observation.
\item Calibration.  To ensure accurate tracking of ionospheric phase
  perturbations, on-source calibration of the amplitude and phase for
  all antennas was executed at the end of each raster scan -- every
  six minutes.  This regimen resulted in an rms phase accuracy
  estimated to be better than 10 degrees.  The highly polarized,
  stable calibration source 3C286 was observed once to establish the
  flux density scale, and to enable calibration of the (R-L) phase
  offset.
\item Polarization: Beam holography requires full Stokes measurements,
  hence all four correlation products (RR, RL, LR, LL) provided by the
  correlator are utilized.  To prevent coupling of the source
  polarization with the antenna polarization, it is necessary to
  utilize an unpolarized target source -- fortunately, the
  polarization of 3C147 is less than 0.1\% at all frequencies between
  1 and 2 GHz.  Separation of the beam polarization (a function of
  position within the beam) from the instrumental polarization (caused
  primarily by the electronics) requires determination of the on-axis
  `D' terms.  These were determined using the on-axis calibration
  observations of 3C147, see below for details.  
Table \ref{tab:param} summarizes the relevant parameters of the observation.
\end{enumerate}

\begin{table}
\caption{Main parameters of the observation. More details can be found in J2018 and references therein.}
\begin{center}
\begin{tabular}{|c|c|}
Array Configuration & `A'\\ \hline
Number of target antennas & 12\\ \hline
Min/Max baselines &  68 m to 36.4 km\\ \hline
Spectral band & L-band (1-2 GHz)\\ \hline
Raster Stepsize  & 309 arcsec\\ \hline
Maximum offset  & 87.6 arcmin \\ \hline
Frequency Resolution & 1000 KHz\\ \hline
Number of spectral channels & 1024\\ \hline
Observed sources & 3C147, 3C286 \\ \hline
Integration Dump Time & 1.0 s\\ \hline
Total Time Duration& 4.1 h\\\hline
Observation Date & 08 June 2014\\ \hline
\end{tabular}
\end{center}
\label{tab:param}
\end{table}%

The observations were taken as part of an overall program to measure
the VLA's primary beam characteristics at all its nine frequency
bands.  These data are public, and are available upon request.  

Calibration followed standard procedures, including removal of
RFI-affected data.  Phase and amplitude calibration was done using the
on-axis observations of 3C147, taken every six minutes.  The amplitude
scale was normalized by dividing the data by the known flux density of
3C147.  3C147 is slightly resolved by the VLA in the `A'
configuration.  The effect of this is trival, as the source extent
(0.7 arcseconds) is tiny compared to the antenna beam size (30
arcminutes) and raster stepsize (4.6 arcminutes).  

Polarization calibration deserves more detailed explanation:  We are
measuring the antenna-based characteristics, while the normal
astronomical procedures are designed to place the observations in the
sky frame.  Converting between the antenna frame (which is that which
the data are taken in) and the sky frame involves adjusting the (R-L)
phases to account for the parallactic rotation of the sky -- a
function of source position.  In AIPS (which was used for the
calibration of the data), this rotation is done after the polarization
calibration is applied to the data -- if no polarization calibration
is applied, the data remain in the antenna frame.  For this work, it
is critical that this phase rotation not be applied since it is the
antenna polarization characteristics -- not the source polarization --
that we are interested in.  A special version of AIPS was generated at
our request to prevent application of the phase rotation upon
application of the polarization calibration.  (For this same reason, it
is important to choose an unpolarized source, which thus bypasses a
complicated adjustment of the antenna polarization characteristics
for the contamination by the source polarization).

Determination of the on-axis (instrumental) cross-polarization
(`D'-terms) was done by two methods: First, utilizing the assumption
that the source is unpolarized; second, by utilizing the parallactic
rotation of the calibrator source during the 4.1 hour duration.
During this period of time, the parallactic angle of 3C147 rotated by
50 degrees, more than adequate for separating the polarizations of
antenna from the source.  The latter method also provides a measure of
the source polarization -- this was less than 0.1\% at all
frequencies.  A comparison of the results for the two methods showed
no measureable difference (to less than 0.05\%) in the D-terms between
the two determinations.  The standard calibration source 3C286 was
observed to enable removal of the (R-L) phase offset between the
oppositely polarized signal channels.

The calibrated data were exported with a special version of the AIPS
program `UVHOL'.  This program recognizes the VLA's holographic modes,
and enables averaging over both target and reference antennas.  For
this paper, averaging was done only over nearby reference antennas,
for each of the individual target antennas.  As noted above, the
parallactic angle correction was disabled for these observations,
ensuring that the data remain in the desired antenna frame.  The
program writes to an external text file, one for each target antenna,
and includes the four complex correlations (RR, RL, LR, LL) for each
of the 35 x 35 = 1225 raster positions.  

These data were then analyzed by special-purpose software, producing
one complex-numbered spectral cube of the scanned beam per antenna per
polarization for all the target antennas under measurement.  The
original data cubes have dimensions of 128$\times$128$\times$1024
spanning over $\sim$1 GHz. Therefore, the beam pattern displays a
spatial scaling factor of $\sim$2 between the minimum and maximum
frequency (Fig. \ref{Fig1-morphology-3}).


\section{Holographic and EM primary beam analysis}
 \label{sec-beamanalysis}
 \subsection{EM model and fitting}
 \label{sec:cassbeam}
VLA antennas have a Cassegrain configuration composed of a parabolic
primary reflector, a hyperbolic sub-reflector and a set of circular
polarization feed horns placed on a ring offset by 1.1 meters from the
antenna axis.  This offset geometry requires the secondary reflector
to be asymmetric to counter the phase perturbations caused by the feed
offset.  This arrangement was selected to simplify band changes (only
the subreflector is rotated -- the feeds remain fixed), but introduces
a squint (separation) in the R and L beams by about 8\% of the FWHM.
The effect on the antenna linear polarization characteristics is very
small.  

The Wiener-Khintchin theorem states that the point spread function
(PSF) of the antenna (i. e. its far-field pattern or beam) is closely
linked to the geometry of the aperture. Therefore, any software
program dedicated to predicting the shape of the beam should take into
account the parameters associated with the geometry of the dish. The
\emph{Cassbeam} PB modelling software was designed with this framework
in mind \citep{Brisken_2003}. \emph{Cassbeam} is a geometric optics
simulator for Cassegrain systems and was designed for analyzing the
VLA and VLBA beam. It produces beam patterns as a function of
direction (in the sky or over the aperture) and also relevant
antenna-related quantities (e. g. system temperature, gain) for each
frequency in a given band.  The \emph{Cassbeam} modelling software
takes eight geometrical input parameters and implement the VLA dish
peculiarities: the location of the feed, the shape of the primary and
secondary and the location of the antenna struts. Diffraction theory
is then applied to derive the far-field pattern of the antenna.
We generated a set of \emph{Cassbeam} beam using the default parameters available for the VLA antenna at the L-band frequencies (see \citet{Brisken_2003}).


 \subsection{2D power distribution}
 \label{sec-2D}
\begin{figure}
\begin{center}
\includegraphics[width=\columnwidth]{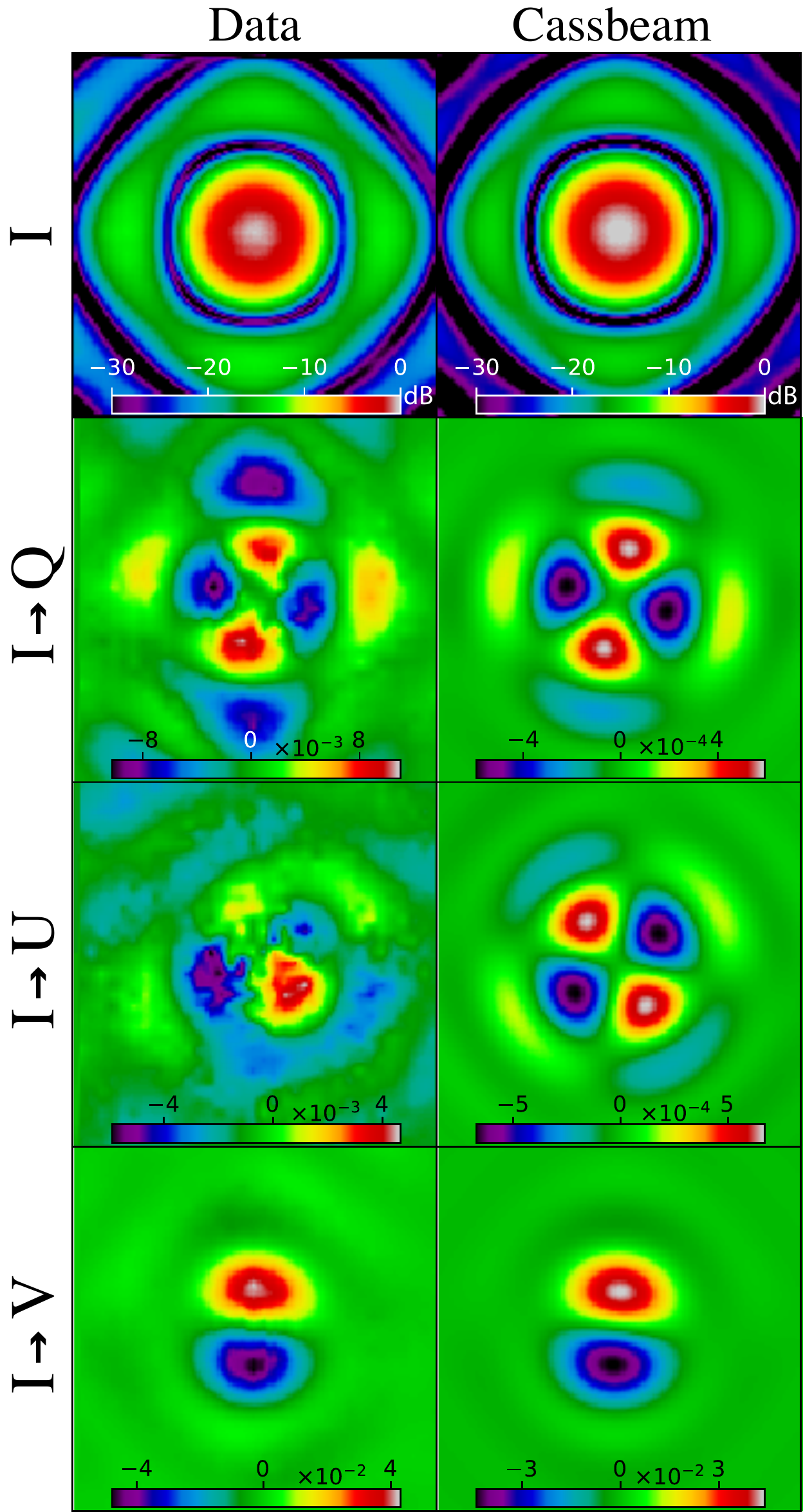}
\caption{Comparison of the morphology of the normalized beam amplitude at 1118 MHz. Each row displays four patterns corresponding to the Mueller matrix element ($I$, $I\rightarrow Q$, $I\rightarrow U$, $I\rightarrow V$) deduced from the holographic measurement of antenna 6 $RR$ $RL$ $LR$ $LL$ beams (left column) and from \emph{Cassbeam} model (right column). Each map was normalized to enhance the visibility of the morphological differences between the terms. However, the color bars encode the amplitude in dB for Stokes I, and the unnormalized amplitudes for the other terms.}
\label{Fig1-morphology-3}
\end{center}
\end{figure}

The typical morphology of the holographic beam (data) is shown alongside that of the predicted EM beam (Cassbeam) at the same frequency in Fig. \ref{Fig1-morphology-3}.
We formed the $2\times 2$ Jones matrix from the Cassbeam and the holography $RR$, $RL$, $LR$, $LL$ elements and we converted it into its $4\times 4$ Mueller matrix form.
Here, we compare each term of the first line of the Mueller matrix ($I$, $I\rightarrow Q$, $I\rightarrow U$, $I\rightarrow V$, where $I\rightarrow Q$ is the second element of the first line of the 4 $\times$ 4 Mueller matrix) at 1118 MHz for the holographic measurement of antenna no. 6 (left column) to the EM model at the same frequency from \emph{Cassbeam} (right column).

This representation is not standard but helps to emphasize the
differences between the expected EM polarized components of the beam
and the measurement. To enable a coarse visual comparison, each
displayed beam has been normalized to its maximum and represented in
dB or in a linear scale. The value of the beam are nonetheless reported on the color bar for reference.
The typical VLA beam shape (seen on the Stokes $I$) can be described as a main circular lobe surrounded by the first side lobe arranged in a 4-fold set of `petals.' 

The overall shape of the holographic primary lobe and first side-lobe are decently reproduced by the EM beam, meaning that the major instrumental effects that could affect the field distribution are correctly modelled by \emph{Cassbeam} (including the known beam squint of the VLA antennas, see e.g. \cite{Chu_1973,Uson_2008,Perley_2016}).
This effect manifests as a slight counterclockwise (resp. clockwise) warping of the $RR$ (resp. $LL$) beam. The original $RR$ and $LL$ holography beams suffer from slight distortion in the shape and distribution of the first and subsequent side lobes.
These distortions in position, side-lobe size, depth of nulls are strongly dependent on the frequency. This can be seen in the two videos available online as supplementary material of the paper. 

For all cross-terms, classically represented as four-fold `petals' for $I\rightarrow Q$ and $I\rightarrow U$ and two-fold `petals' for $I\rightarrow V$, we can match the main features between holography measurements and the model, in particular for $I\rightarrow Q$ and $I\rightarrow V$ displaying a relatively good agreement between the two. However, the  $I\rightarrow U$ appear to be weaker than the $I\rightarrow Q$ pattern. Such a difference might be explained by the effect of the antenna legs, oriented with $Q$ but not with $U$ which can combine with the effect of having an asymmetric sub-reflector. The overall rotation of the cross-terms are correctly reproduced by the model and is associated to the particular VLA antenna geometry. The distortion and apparent noise on the holography maps are due to the combination of the data quality and the measurement principle. 
As VLA antennas have an alt-az mount, and as we are observing an unpolarized source in the holography measurements process, no specific derotation is required to account for parallactic angle rotation (see Sec. \ref{sec-holovla}).
However, this issue, visible in the $I\rightarrow U$, does not invalidate the methods and results presented in this study.


For all terms, inspection of the spectral morphology shows an overall beating effect with increasing frequency. While the scale of the whole beam slowly decreases with frequency (following a $1/\nu$ law), the location of the beam centre, as well as the size of the beam and size and distribution of the sidelobes, beat with frequency. This is due to the blockage of the primary reflector by the sub-reflector of the antenna, which probably allows for a standing wave to form and induce a frequency beating. 

This first coarse comparison of the Mueller terms leads us to the conclusion that EM simulations with \emph{Cassbeam} represent the overall distribution of the power in the measured beam, accounting for some instrumental effects up to the secondary nulls. In the following, we will focus on the $RR$, $RL$, $LR$, $LL$ representations of the measured and simulated quantities.
The holographic beam, relying on the observation of two kinds of point sources (an unpolarized and a polarized one), shows a high degree of distortion, especially in the cross-terms components, which can be caused by the quality of polarization calibration, a quality known to vary with frequency. Theses distortions induce a shift of the effective centre of the main lobe, causing pointing offset errors (see Sect. \ref{sec-pointing}).

\begin{figure*}
\begin{center}
\includegraphics[width=\linewidth]{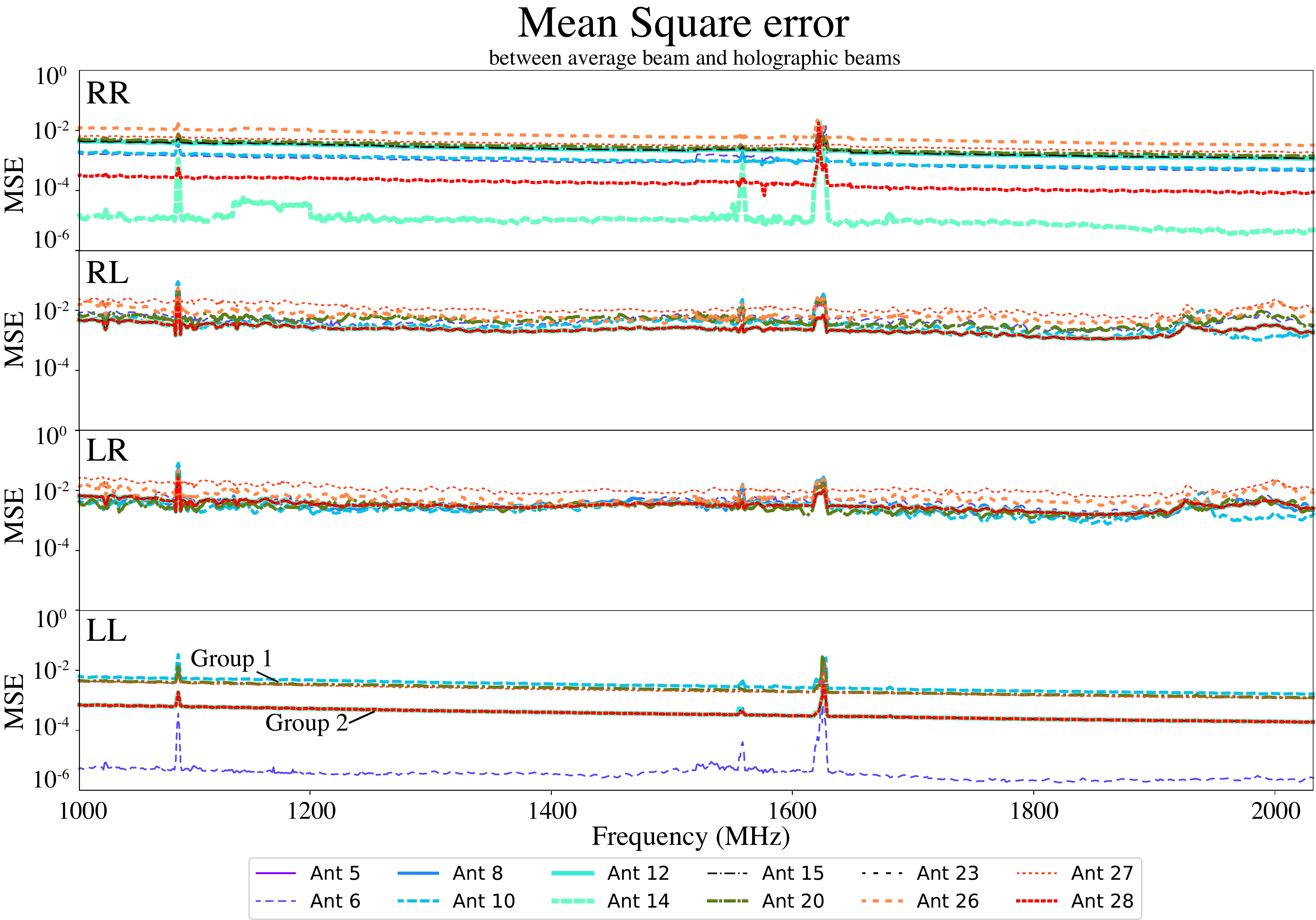}
\caption{Evolution of the Mean Square Error between each of the 12 holography beams and the averaged holography beam as a function of frequency, for each antenna, for each correlation. Group 1 refers to antenna no. 10, 20, 27, group 2 to Antenna no. 5, 8, 12, 14, 15, 26, 28 }
\label{Fig2-MSE}
\end{center}
\end{figure*}

To quantify how all the \textit{target} antennas behave, we computed, at each frequency and at each polarization, the normalized average beam over the 12 target antennas. We can also quantify the dispersion of the beam of the different antennas w. r. t. this normalized average beam. In order to do that, we computed the \textit{average} holographic beam over all the 1024 frequency channels. Then, at each frequency channel, we computed the Mean Square Error (MSE) between the average beam and each of the normalized beam. In a classical context, the MSE is used as an evaluation of the quality (e. g. fidelity) of an image w. r. t. to a reference image. Here, we used the MSE as a quantity that serves to evaluate the dispersion of each antenna around the average behaviour as a function of frequency for each of the correlations.
We show the variation of this MSE in Fig. \ref{Fig2-MSE} on a logarithmic y-axis.

We note that for $RR$ and $LL$, the variation of the logarithmic MSE with frequency for each antenna follows a smooth linear trend associated with a decreasing power law. This suggests that the antennas are close to the average behaviour at higher frequencies.
The overall MSE varies between 0.01 -- 1 per cent for $RR$ and $LL$ and around 1 per cent for $RL$ and $LR$.

In $LL$, we notice that the 12 antennas are split in two groups, group 1 (Ant 10, 20, 27, the latter two coincide) around $5\times10^{-3}$ to $8\times10^{-3}$, group 2 (Ant 5, 8, 12, 14, 15, 26, 28) around $10^{-3}$ displaying the same coinciding behaviour that follows the average trend and an isolated antenna (Ant 6) around 8$\times 10^{-6}$.
There is no obvious correlation between the arbitrary composition of each group and a common physical localization of the target antennas in the array.
In $RR$, the dispersion of the behaviours of different antennas is higher, ranging from $10^{-5}$ to $\sim$$10^{-2}$, making any antenna grouping irrelevant. It is interesting to note at this point that with the same figure of merit, the results we obtain are not symmetrical between $RR$ and $LL$.

Conversely, for $RL$ and $LR$, the dispersion of the MSE is more compact, ranging around $10^{-2}$ on average, (except for Ant 26 and 27) from one antenna to another but display a noisier pattern across frequency.

This plot shows the general behaviour of the distortions of the different holographic beams w. r. t. the average beam.  The intrinsic variations of the holographic beam will be described in the sections that follow. 

The coarse visual inspection of the holographic beams, and the quantification of their variation w. r. t. the average beam, are necessary as the first step towards a better understanding and characterization of the relevant fluctuations.
A good knowledge of the variation of the beam morphology in frequency is needed for HDR calibration and imaging. Variation at the percent level can have a relatively strong impact on the quality of the calibration of offset weak radio sources that can be polluted by an insufficiently modelled (far) sidelobe of the beam.
%
%
There is scope for further parametric beam modelling as in \cite{Jagannathan_2017_2} (referred to as J2018 in the following). However, a deeper phenomenological modelling of the beam is necessary to represent the beam and use it for calibration in a practical way. In the next section, we will show how we can unfold the different contributions of the beam fluctuations and represent them with the smallest number of coefficients while keeping a high fidelity of representation.

 \subsection{Effective FWHM as a function of frequency}
 \label{sec-fwhm}
 
 In order to inspect the various contributions accounting for, on the one hand, the distortion of the beam w. r. t. EM simulations, and on the other hand, the dispersion between antennas, we start by fitting a two dimensional elliptical Gaussian to the main lobe. This will enable us to investigate the variations of the FWHM and pointing offset errors. As it is non-trivial to determine the centre of the $LR$ and $RL$ beams, we restricted the study to the $RR$ and $LL$ beams, whose variation with frequency is assumed to be representative of that of the $RL$ and $LR$ beams.
We used the \emph{optimize.leastsq} procedure from \textsc{Scipy} \citep{scipy} to perform a least-square optimization to find the optimal elliptical Gaussian that fits the primary lobe at each frequency. The residual map comes as a by-product of the optimization, giving the difference between the data and the fitted model. We used the variance of the residual map to derive a goodness of fit. 

\begin{figure}
\begin{center}
\includegraphics[width=\columnwidth]{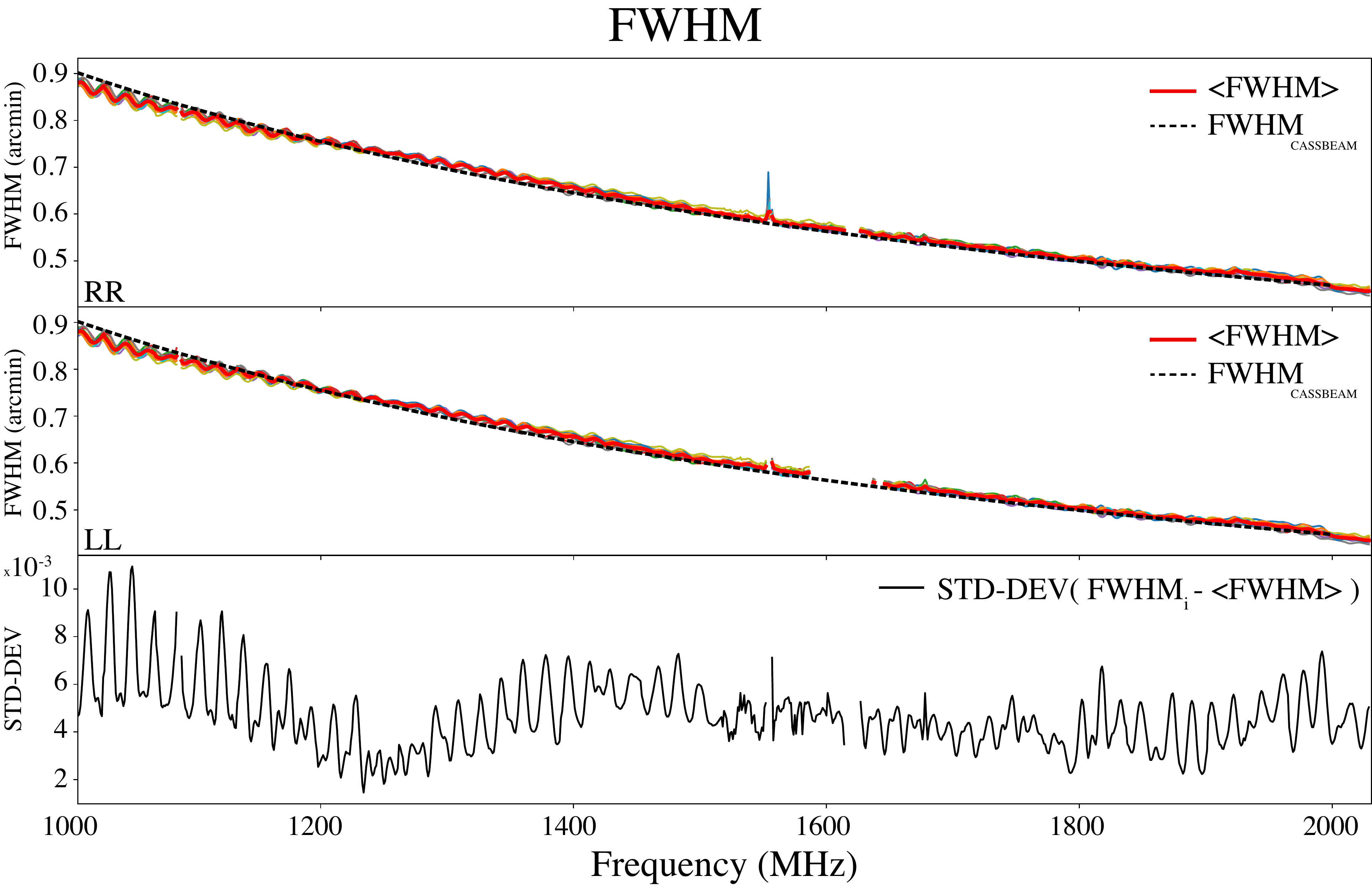}
\caption{Evolution of the fitted FWHM as a function of frequency for the $RR$ and $LL$ beam amplitude. The average FWHM variation is plotted with the red line. The dash line represent the FWHM derived from the fit of the \emph{Cassbeam} model. (Bottom row) Standard deviation across the 12 antennas after removing the average trend.}
\label{Fig3-FWHM}
\end{center}
\end{figure}

We display, in Fig. \ref{Fig3-FWHM}, the evolution of the FWHM fit parameter as a function of frequency for all the target antennas as well as that of the average beam taken over the 12 target antennas.
Blank data represent the location of high variance channels associated with RFI and distorted beam shapes.
The FWHM in the $l$ and the $m$ directions can be considered equal, due to the overall symmetry of the beam. This confirmed by the fact that the major and minor axes of the fits were very close over the L-band. So we consider the beam effectively circular to derived the FWHM values. As a consequence, the position angle information is therefore not useful in our study. In addition, the covariance of the size fits at each frequency was too small to be represented on the plot. 

Several effects can be noticed in order of decreasing importance: i) the scaling effect due to the frequency change, ii) the low-level oscillations as a function of frequency and iii) the possible variation of the amplitude of these oscillations. 
First, the variation of the FWHM of the average holographic primary lobe decreases, at first order, as a function of $1/\nu$. We can, however, note that a more complex relationship is required to account for the real frequency dependence. The dashed line represents the expected $1/\nu$ trend derived from the \emph{Cassbeam} model. One can therefore clearly see how the measured frequency trend deviates from the theoretical one. In particular, at the low-end of the frequency windows which is relatively free of RFI, the deviation can be due to fine instrumental effect. For example, the VLA’s ortho-mode transducer (OMT) and the horn were not designed for good characteristics below 1.2 GHz. In particular, the horn is significantly undersized (in order to allow the seven other feeds to fit the focus ring), so the sub-reflector is significantly over-illuminated below 1.2 GHz. This can be potentially associated with the larger deviation at low frequencies.
The second effect is related to the presence of a standing wave between the primary and secondary reflector of the antenna. It induces a strong frequency ripple on the whole beam pattern. This effect was already described by previous authors (J2018 and references therein) and manifests at an average frequency of 17.2 MHz (which corresponds to twice the light path travelled by the standing wave). The third effect is the possible attenuation of the ripple amplitude as the frequency increases.

J2018 addressed all these effects by relying on the fitting of the physical model parameters around their default design values for each frequency channel. This way, any deviation from the $1/\nu$ trend could be absorbed into the fit and can be associated with a mechanical interpretation.

\subsection{Squint \& Pointing error}
\label{sec-pointing}
 
\begin{figure}
\begin{center}
\includegraphics[width=\columnwidth]{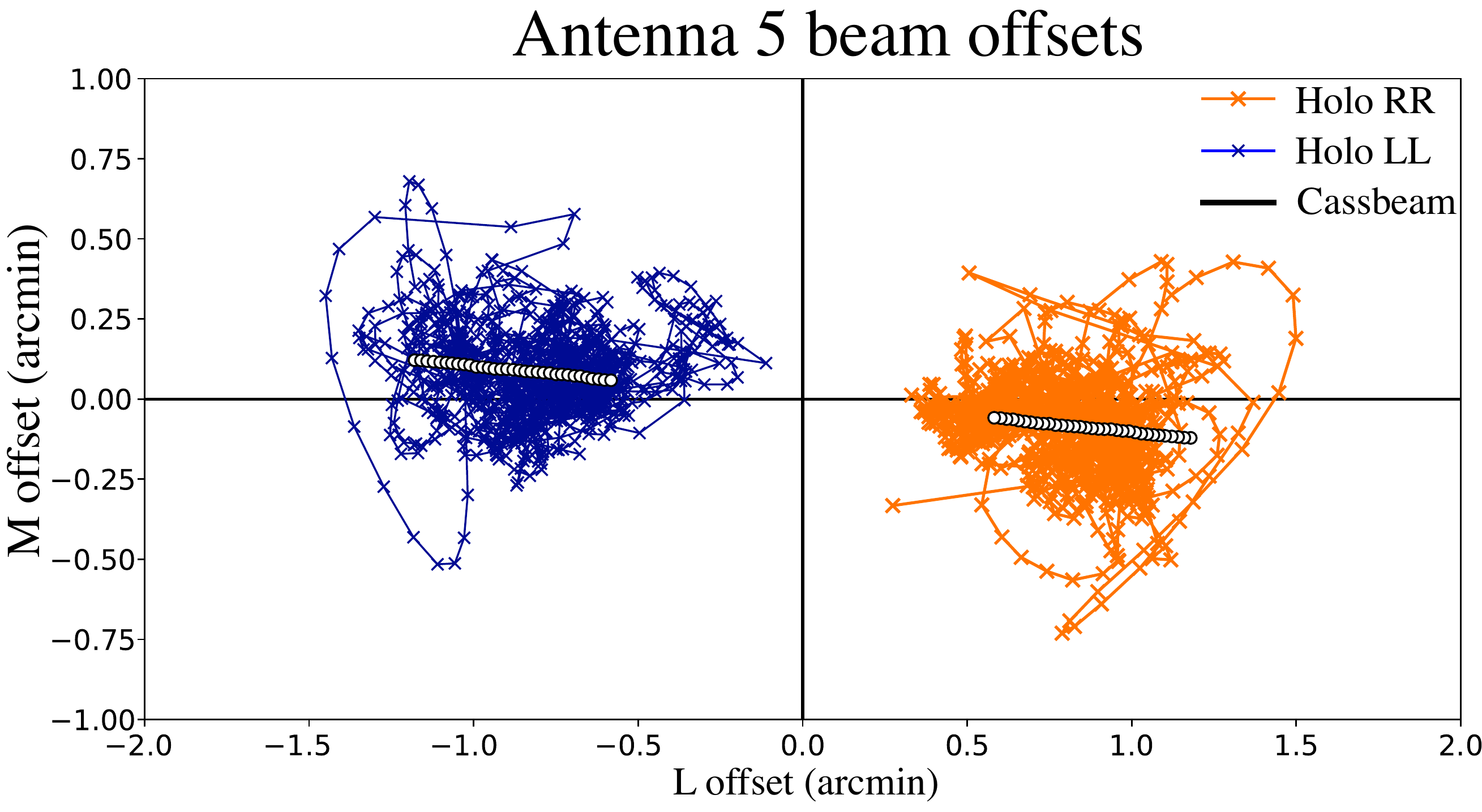}
\caption{2D Spatial distribution of the pointing offsets with frequency for the $RR$ (left) and $LL$ (right) beam compared with the expected pointing offset from \emph{Cassbeam}.}
\label{Fig4-offsetone}
\end{center}
\end{figure}

 We use the same fitting procedure as in Sect. \ref{sec-fwhm} to solve for the antenna pointing offset at all frequencies. Fig. \ref{Fig4-offsetone} shows the distribution of the $l$ and $m$ offsets as a function of frequency, represented on the same $l$,$m$ plane. The angular tilt of the holography and the corresponding EM pointing offset suggest that it is of instrumental origin.
 The two linear tracks correspond to the \emph{Cassbeam} expected pointing offset induced by the beam squint which is incorporated in the physical model of the antenna. Similarly, the error bars are too small to be represented on the figure.
 
\begin{figure}
\begin{center}
\includegraphics[width=\columnwidth]{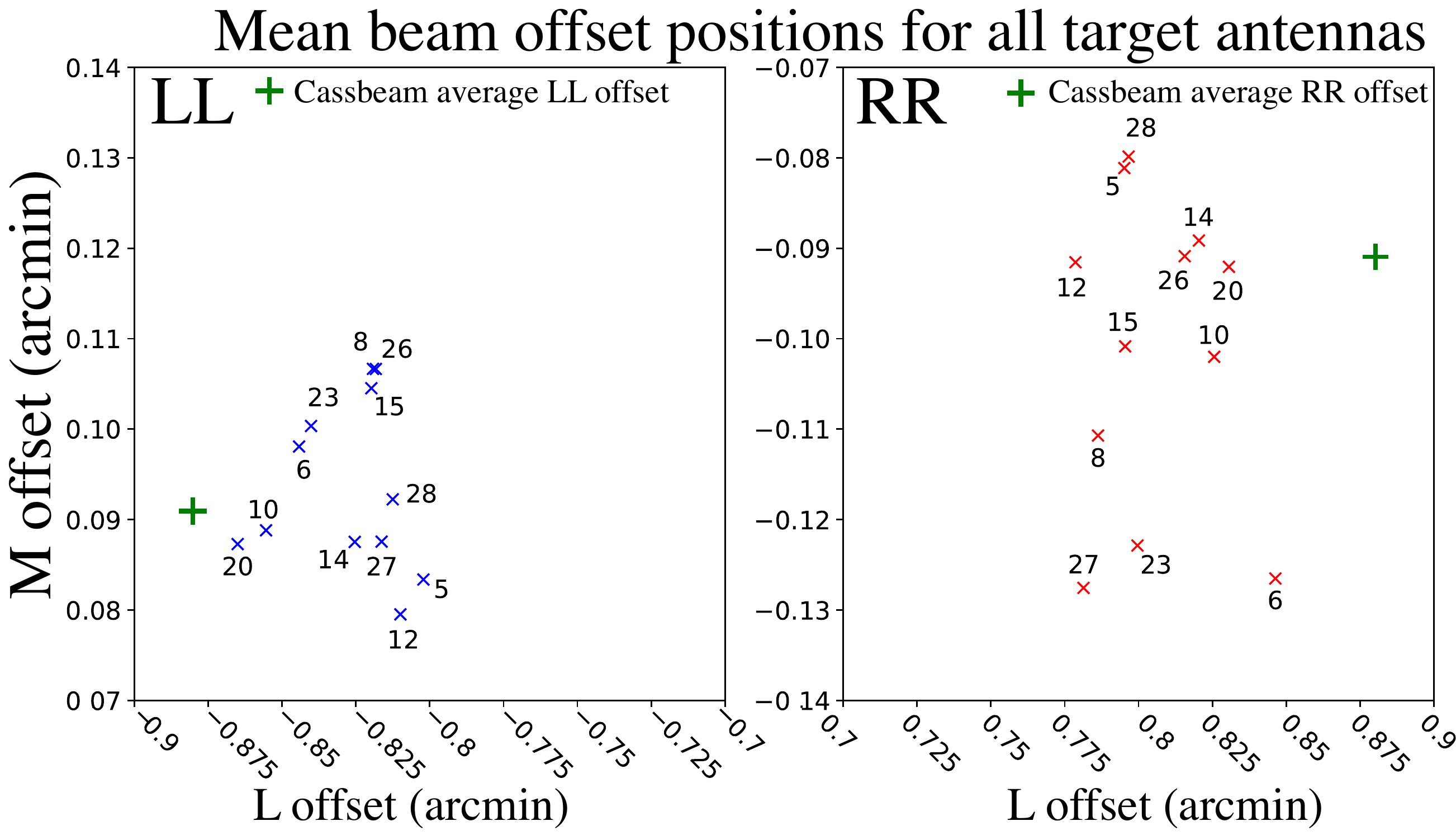}
\caption{2D Spatial distribution of the average pointing offset, for each antenna and each polarization. The cross marks the expected average location of the \emph{Cassbeam} model.}
\label{Fig5-offsetall}
\end{center}
\end{figure}

For each antenna, we took the frequency-averaged beam offset position for $RR$ and $LL$ beams. Results are represented in Fig. \ref{Fig5-offsetall} which highlights a systematic offset and a spreading of effective holography pointing error compared to the \emph{Cassbeam} expected pointing offset. Average and dispersion of the pointing offsets for $RR$ is $\delta l=48.4 \pm 1.2$ arcsec, $\delta m=-6.1 \pm 0.98$ arcsec and $\delta l=-49.6 \pm 1.1$ arcsec, $\delta m=5.6 \pm 0.5$ arcsec for $LL$. The reference average \emph{Cassbeam} pointing offset is $\delta l=52.8$ arcsec, $\delta m=-5.5$ arcsec for $RR$ and $\delta l=-5.5$ arcsec, $\delta m=52.8$ arcsec for $LL$.

Again, there is no obvious relationship between the location of the antenna on the ground with the amplitude of the mean $RR$/$LL$ offset per antenna. There is also no possible comparison with the empirical grouping we noticed in Fig. \ref{Fig2-MSE}, suggesting that the variation of the offset position has a random nature. However, we do note, in Fig. \ref{Fig5-offsetall}, that a systematic shift of the set of all mean pointing offset and that given by \emph{Cassbeam}, suggesting that there is still a residual shift between the effective beam maximum and the theoretical squinted location of the $RR$/$LL$ beam maxima.

\begin{figure}
\begin{center}
\includegraphics[width=\columnwidth]{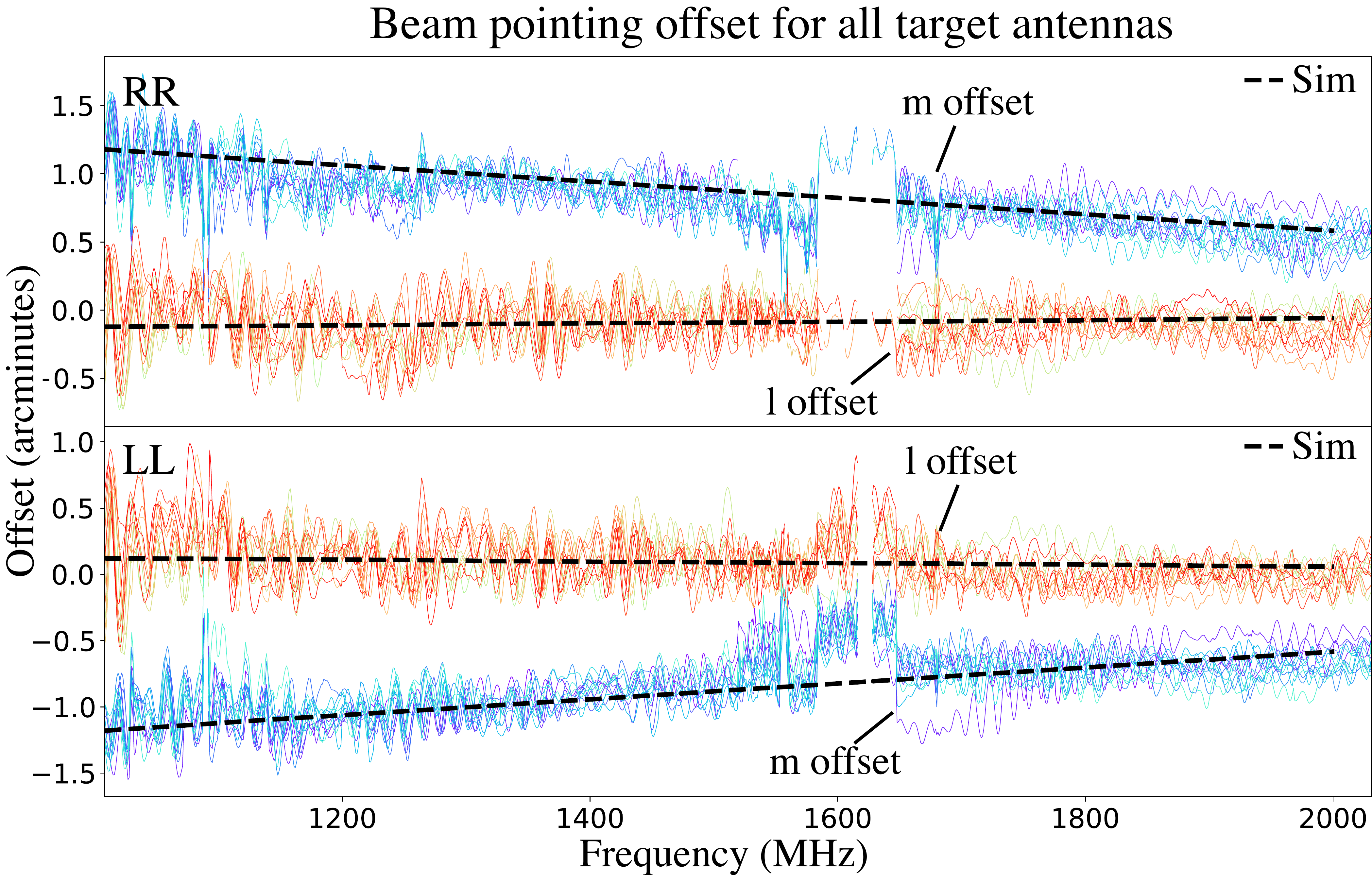}
\caption{Spectral distribution of the $l$ and $m$ pointing offsets, for all 12 antennas and the $RR$ and $LL$ polarization.}
\label{Fig6-offsetfreq}
\end{center}
\end{figure}

Pointing errors can lead to biased estimation of the flux density of sources. If a source is supposed to be located in the first null ring of the beam but falls just near this null, the data will contain some flux contributed by this source, whereas the sky model combined with a wrong beam model will lead to poor fitting, e.g. especially for faint sources in a >10$^6$:1 dynamic range image.
It is, however, possible, to artificially match the pointing offset with the use of the physical parameter of the EM antenna model.
 Fig. \ref{Fig6-offsetfreq} shows the distribution of the $l$ and $m$ offset as a function of frequency, for each antenna.

Fitting for the FWHM and the pointing offset error at each frequency provides a way to correct for these two effects by using a restricted set of linear transformations derived from the fitted parameters. Once the FWHM and pointing offset error has been taken out, we generated a set of scaled and shifted holography beam. It is then possible to investigate the beam pattern modelling on a deeper level.

\section{Search for sparse beam representations}
\label{sec-beammodelling}
\subsection{Sparse representations}
\label{sec-sparsity}
Signal processing has made a huge leap forward thanks to the mathematical framework of sparse representations, convex optimization and, more generally, the compressed sensing framework \citep{CS} and data compression.
Searching for sparse representations is a powerful approach to understand the nature and distribution of \textit{energy} in a signal. It can be said, based on Occam's razor, that a signal is efficiently represented in some space if the support that describes this signal is small, i. e. represented by a small number of representative coefficients. An obvious example of this is to consider a time-dependent cosine function of frequency $\nu$. In the signal domain, it requires a lot of coefficients (with a minimum sampling rate of the Nyquist frequency $2v$) to account for the sinusoidal variation in time without any prior knowledge on the signal. However, we do have a prior knowledge of the nature of the signal. Indeed, in the Fourier domain, our signal only requires two complex coefficients, represented by two Dirac delta functions. Here, the associated Fourier kernel basis is said to be a sparsifying basis for the signal, i. e. the signal is sparse.\footnote{A trivial sparse dictionary representing the data, can be the data itself, but such representation does not add any additional information w.r.t. the data, and does not help to model a slightly different data set.} 

In our study, we want to understand and model the variations of the beam by searching for a set of decomposition of the signal in sparse bases. Each holography measurement is a 3D data set which contains a series of 2D spatial frames, distributed across frequency for each cross- and co-polarization.
If such sparsifying bases exist, they will help to compress the features and encode the spatial and spectral variations of the beam in a small set of relevant coefficients. The motivation for this compression is to determine a set of empirical degrees of freedom that will, in turn, become `new' unknown parameters to be incorporated and solved for during the direction dependent (DD) calibration process.

There are two main ways to find suitable sparsifying bases: from bases derived from the data itself (data-based representations, constructed with Principal Component Analysis (PCA), Singular Value Decomposition (SVD) or dictionary learning techniques) or from pre-existing bases that provide good \emph{a priori} description of the signal (Fourier components, wavelets, Zernike modes, spherical harmonics to name a few).

Recent work on beam modelling relies on the decomposition of the beam on Characteristics Basis Function Pattern \citep{CBFP1,CBFP2}, which exploits the PCA method. We will also start with this method by trying to study both spatial and spectral variation of the holographic beam.

\subsection{A modified Cassbeam model}
\begin{figure}
\begin{center}
\includegraphics[width=\linewidth]{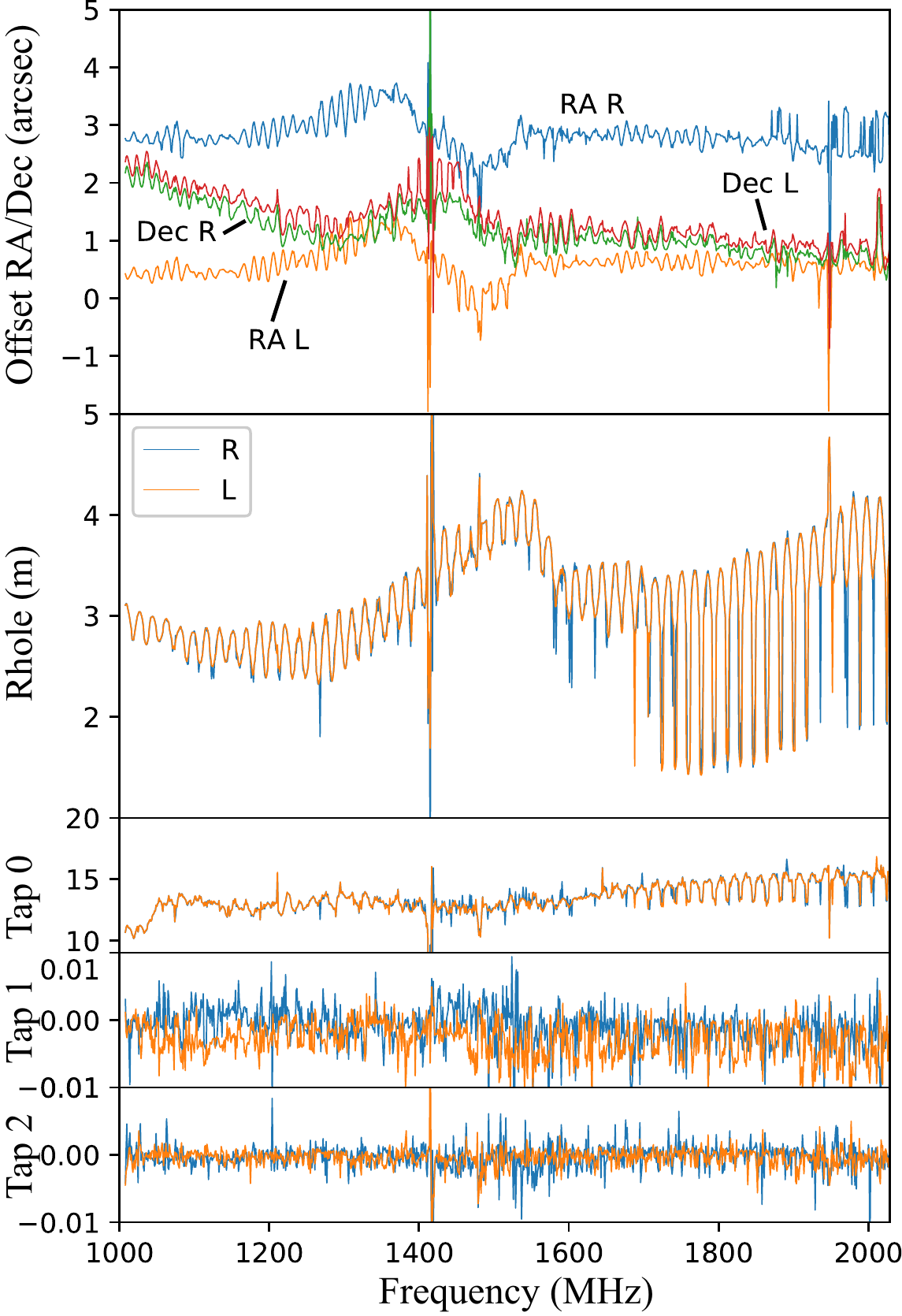}

\caption{Fitted Cassbeam model from J2018 showing the frequency dependency of the pointing offset (Offset X, Offset Y) measured after ray tracing with Cassbeam model, the radius of the central hole (Rhole) and the aperture illumination taper coefficients (Tap 0, Tap 1, Tap 2) for each R and L feed.}
\label{Fig7-Preshantparam}
\end{center}
\end{figure}

%
%
 J2018 developed a fitting code wrapping the \emph{Cassbeam} parametric model to fit the observed frequency behaviour of the holographic beam. By performing this fitting at each frequency, they derive a series of geometric parameters accounting for the rippling effect that was discussed earlier. Fig. \ref{Fig7-Preshantparam} shows the frequency variation of geometric coefficients for antenna no. 6 for both R and L feed. The fitted parameters are the feed offset in ($l$,$m$) (i.e. Offset X, Offset Y), the radius of the central hole (i.e. Rhole) and the polynomial that fits the aperture illumination taper function (Taper-0, Taper-1, Taper-2).
 
The ripple effect is absorbed mainly in the fit of both feed offsets and particularly in that of the central hole radius. J2018 states that they can model the central blockage as an apparent blockage by capturing this frequency variation, presumably associated with the effect of standing waves.
In total, $7 \times N_{\textrm{chan}}$ coefficients per feed and per antenna are required to reproduce the observed holography beam with reasonable accuracy (see J2018).

In our study, we reproduced and extend part of the results obtained in J2018 and by performing a statistical analysis of all antennas PB behaviour, and we look for sparse representation of the spatial and spectral variation of the beam.

\subsection{Data-driven representation}
\label{sec-PCA}

Holography measurements were originally developed to investigate the surface quality of an aperture. The distribution of amplitude and phase over the aperture can be associated with surface errors that can then be physically adjusted on the antenna.
By taking the inverse Fourier transform of the holography beams, we can represent the power distribution over the aperture. Since all physical information about aperture illumination should, by definition, lie inside the aperture, one can filter out all remaining information outside of it. By performing a Hard Thresholding (HT) operation (nulling the low value pixels mostly present outside the synthetic aperture domain, \cite{Donoho_1994}), one restricts the measurements to the aperture plane, thus eliminating distortions. This HT step is equivalent to a denoising step, resulting in a smoothing of the beam. The residuals of the HT denoised holography beam are shown in Fig. \ref{Fig14-comparisonresiduals}.

Principal Component Analysis (PCA) is a well-known data decomposition method, easily computed through a Singular Value Decomposition (SVD). This method helps construct a basis of vectors (associated to eigenvalues) that matches the main features of a data set. Starting from a signal living in an N-dimensional space, one can represent the data as a collection of data points lying in the N-space and possibly forming a cloud of complex morphology. The motivation for this method is to search for a lower dimensional subspace composed of a minimal number of basis vectors that explain most variations of the data set. These `principal' vectors are associated with the axes that sparsely describe the morphology of the cloud in N-space. This means that a correct set of axes enables the computation of the coordinates of any of the data points in a sparse way, thus minimizing the distance. This method naturally enforces sparse representations of the data set, along with providing a basis of eigenvectors (called the principal components) and the associated eigenvalues.

Once these basis vectors have been found, the new coordinates of the data points can be computed in the basis and one can perform dimensionality reduction similar to data compression. Most coefficients that are generated from the projection of the data in the subspace will be concentrated in a few major principal components. As the energy of the signal concentrates in a few main coefficients, we can start producing low-rank approximations of the original data that still reproduce the morphology and the variation of the beam in a robust way.

Classical PCA requires the data to be centred and normalized, particularly to ensure the robustness of the method against noise. In our study, we did not do this by taking out the average beam, because the noise level is small in our data set.
For each antenna, we have a 3D holography cube (2 spatial axes x 1 spectral axis). Each 2D frame is converted into a 1D vector, forming a rectangular matrix $D$ of dimension $N_{\mathrm{pix}}^2$ $\times$ $N_\text{freqs}$. 
We cannot search for a singular value decomposition of this matrix a priori since it is not a square matrix. Instead, PCA enables to search for the eigenvectors and eigenvalues of the covariance matrix of $D$, Cov($D$) that can reach a high dimension $N^2_{\text{freqs}}$. This covariance matrix is often costly to compute and constitutes the only strong limitation of the method as the data set becomes large. Computing Cov($D$) also leads to loss of precision.
However, PCA and SVD are closely tied to one another, and it is possible to use algorithmic shortcuts that avoid computing the full covariance matrix. 
%
%

\begin{figure}
\begin{center}
\includegraphics[width=\columnwidth]{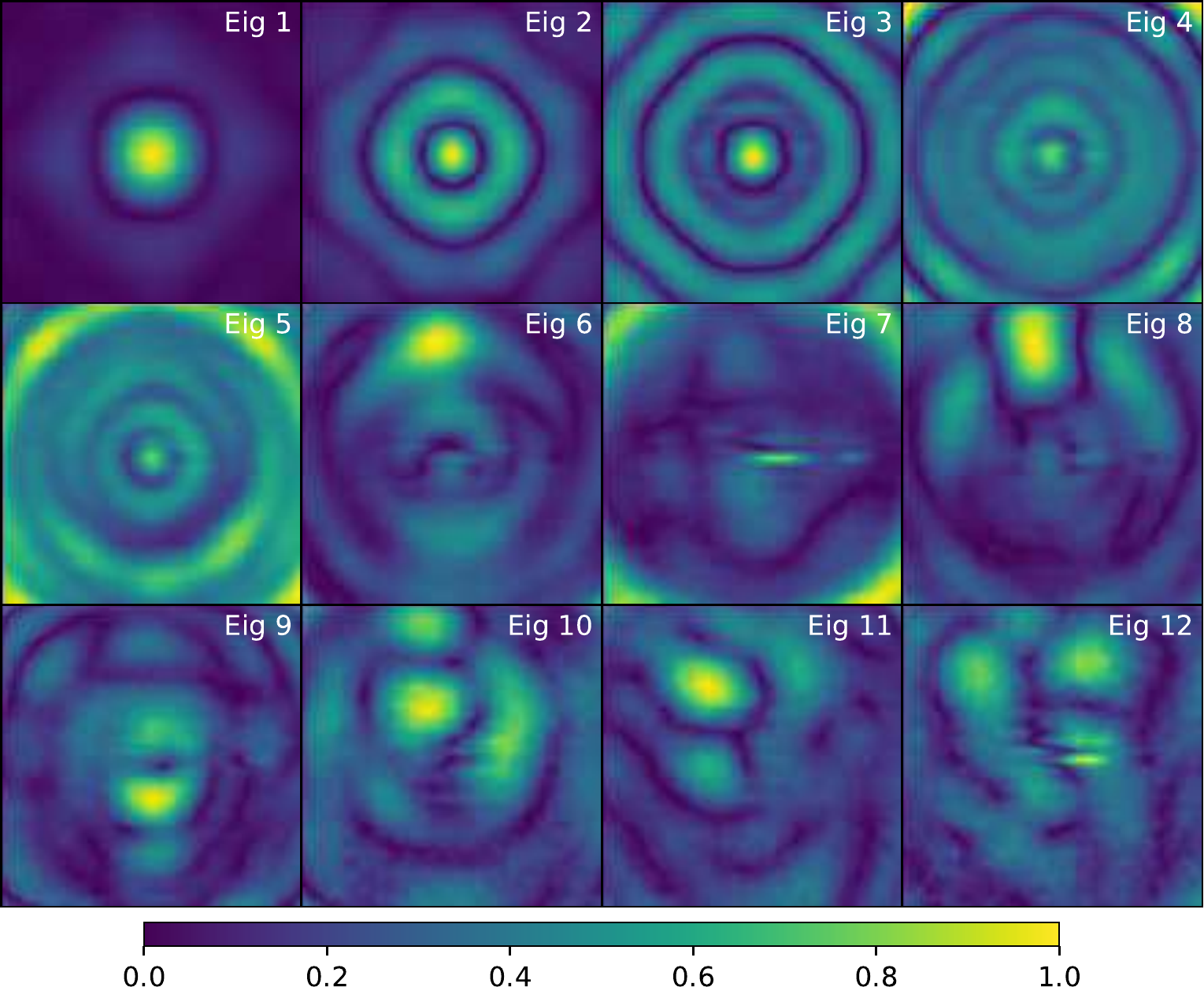}
\caption{First 12 dominant PCA modes for antenna 6 that were ranked among the 5$^{\text{th}}$ across the L-band. Each subplot represents a normalized PCA mode.}
\label{Fig8-SVDant6vector}
\end{center}
\end{figure}

\begin{figure}
\begin{center}
\includegraphics[width=\columnwidth]{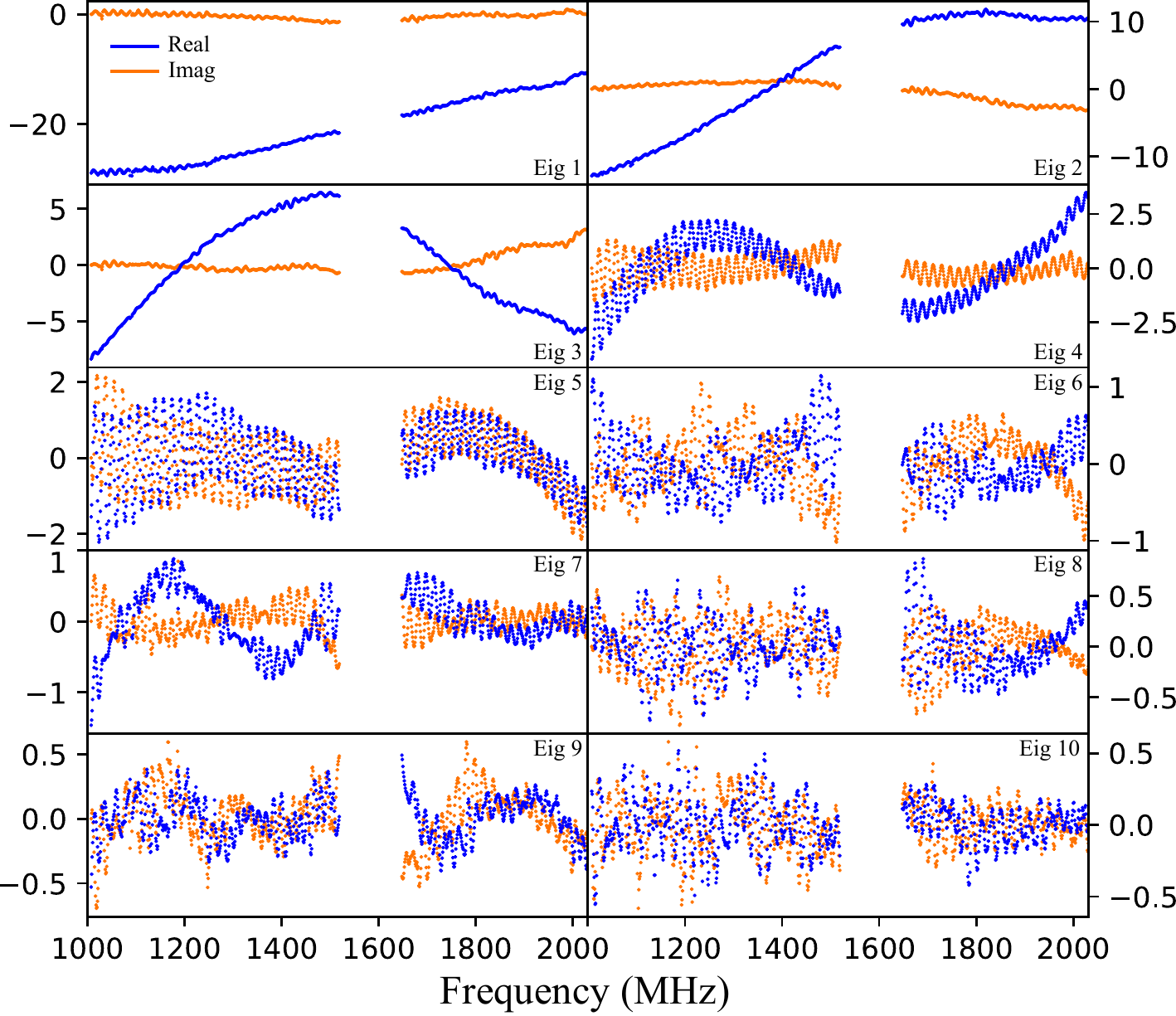}
\caption{Spectral behaviour of the first 10 SVD/PCA components (real and imaginary) for antenna 6.}
\label{Fig9-SVDant6coeff}
\end{center}
\end{figure}

We performed a PCA on each antenna's complex-valued holography beam and we derived the corresponding eigenvectors (or \textit{eigenbeams}) and eigenvalues. 
Fig. \ref{Fig8-SVDant6vector} presents the first 12 main eigenbeams representing the data over the L-band. The PCA/SVD provides the most dominant eigenvectors and eigenvalues ranked by decreasing power. Therefore, it is fairly easy to obtain the low-rank approximation corresponding to a desired precision and fidelity.
The most powerful eigenbeam contributes to model the primary lobe at low-frequency and the four next eigenbeams contribute to the side lobes and to the primary lobe at higher frequencies. Higher degrees of decomposition corresponds to the deviation from the theoretical shape of the beam. The main features of the beam can be reproduced, over the L-band, using only a few characteristics beams.
We can then proceed to dimensionality reduction by keeping only the first $k$ components that mostly represent the data. The beam $b$ will therefore be approximated by Eq. \ref{svd}:
\begin{equation}
\tilde{b}=\sum\limits_k u_k w_k v_k^T
\label{svd}
\end{equation} 
where $u$, $v$ are the complete unitary matrices and $w$, the complete 
diagonal matrix produced by the SVD of matrix b. The index $k$ refers to the $k\times k$ sub-matrix.
Fig. \ref{Fig9-SVDant6coeff} presents the spectral behaviour of some of the complex coefficients (the real part and imaginary part are shown rather than amplitude and phase). In order to enhance the readability of each decomposition order, each curve is represented with its own scale represented on the left or on the right of each plot. One can see that the amplitude of the coefficients decreases with the degree of decomposition $k$, suggesting that low-rank approximation can be done without severe loss of information.
The first 3 components are mainly dominated by a smooth variation in frequency in addition to a small ripple component. The amplitude of the ripple in the first three orders has a peak-to-peak amplitude of $\sim$1.3 arb. units. 
Starting from the 4$^{\text{th}}$ onwards, the coefficients are mostly dominated by the ripple at $\sim$17 MHz in addition to a small smooth trend in a noisy background. The ripple reaches a maximum peak-to-peak amplitude of $\sim$3 arb. units in the 5$^{\text{th}}$ order before decreasing in larger orders.



In general situations, PCA/SVD is dependent on the data, but is independent of the \textit{morphological} content of the data. To illustrate this fact, we can look at the face recognition studies using PCA and which is analogue to beam modelling (see e.g. \cite{PCAEigenvectors}). In face recognition, the PCA produces `eigenfaces' that do not encode separately the various features composing a human face (e.g. eyes, mouth, etc.) but spawns an adequate basis composed of mixed features. The eigenbeams are not beams, but the mixing of the different beam features (primary lobes, side lobes), forming a relevant basis to represent all beams of the data set. In combination with a set of weight, also determined by PCA, the basis enables the data set to be represented by a small set of coefficients forming a sparse representation.

We can note that all antennas share similar spectral trend in their coefficients, making their description redundant, i.e. potentially modelled by a single model with high compressibility. For the sake of modelling each antenna, we can think of deriving a single common set of eigenbeams that could represent all the beams. Given a common basis, we can derive new coefficients by computing the dot product between each antenna beam and the eigenbeams. 
In the next section, we will select another approach which relies on finding a decomposition of the beam at all frequencies based on its structure.


\subsection{Orthonormal basis representation}
\label{sec-zernike}
Aperture illumination of an approximate circular antenna can be approached with the tools of optical physics. The morphology of the $RR$/$LL$ primary beam invites us to use an orthogonal basis that can grasp the various features and distortions of the beam, namely the Zernike polynomials. This orthogonal basis is designed to decompose into modes defined on circular apertures, which will distribute the energy of a signal into specific orthogonal features. 
Zernike polynomials are widely used in adaptive optics to sparsely represent the optical deformation of a telescope's primary mirror with piston, tip-tilt, defocus and other terms (see for example \cite{zernike}).
Zernike modes are particularly suitable for modelling a signal with very few modes, e.g. the characteristic VLA primary lobe and four-fold petals of the first side lobes. For our study, we started from the original holography cubes which do not include FWHM scaling with frequency or the beam pointing offset correction. Indeed, the pointing offset can be fully encoded in the first tip-tilt modes.

\begin{figure}
\begin{center}
\includegraphics[width=\columnwidth]{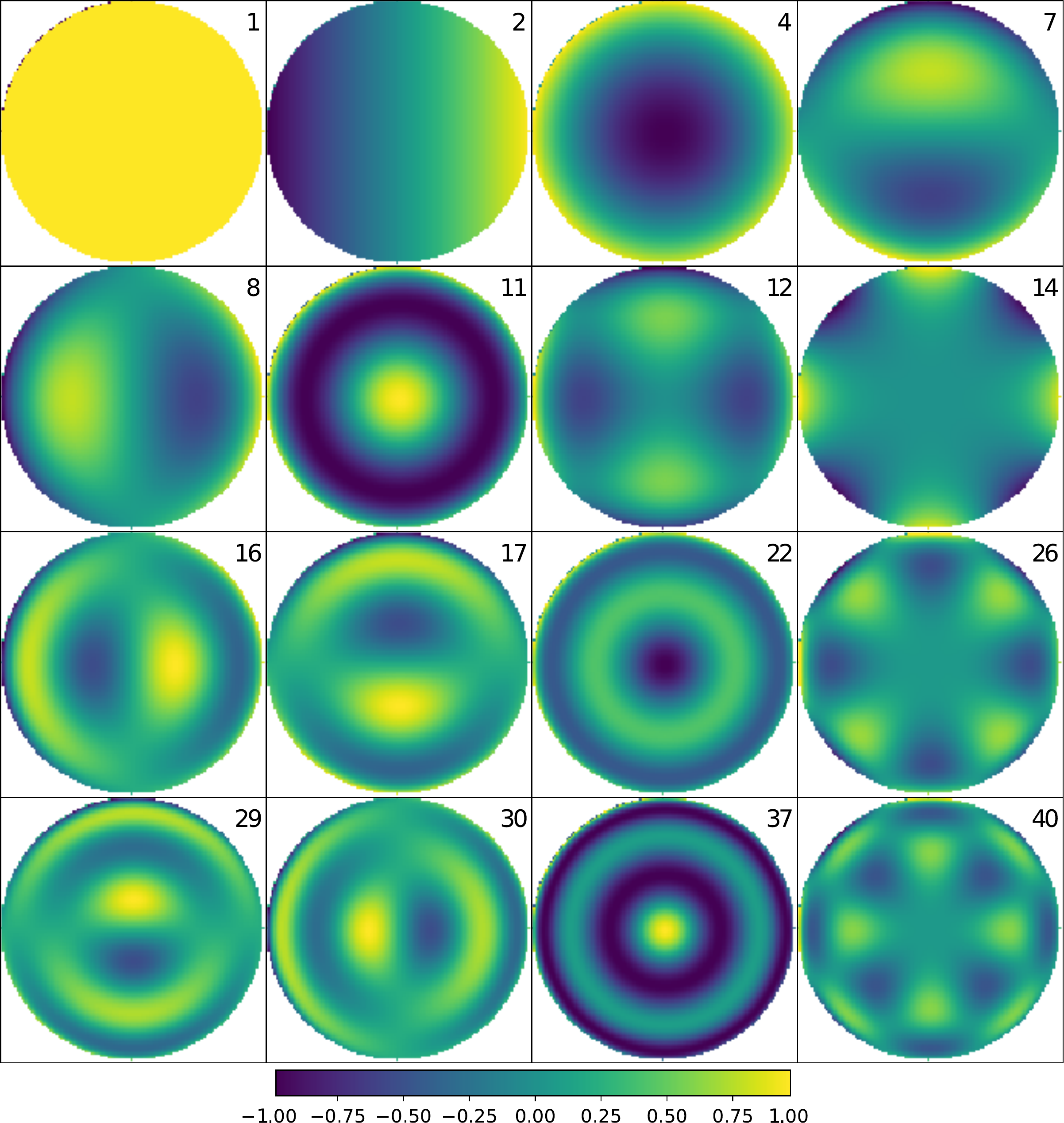}
\caption{First 16 dominant Zernike modes for antenna 6 that were ranked among the 20 across the L-band. Upper right number represents the Noll index associated with the Zernike polynomials. Each subplot represents a normalized Zernike mode.}
\label{Fig10-ZERant6vectornew}
\end{center}
\end{figure}

\begin{figure*}
\begin{center}
\includegraphics[width=2\columnwidth]{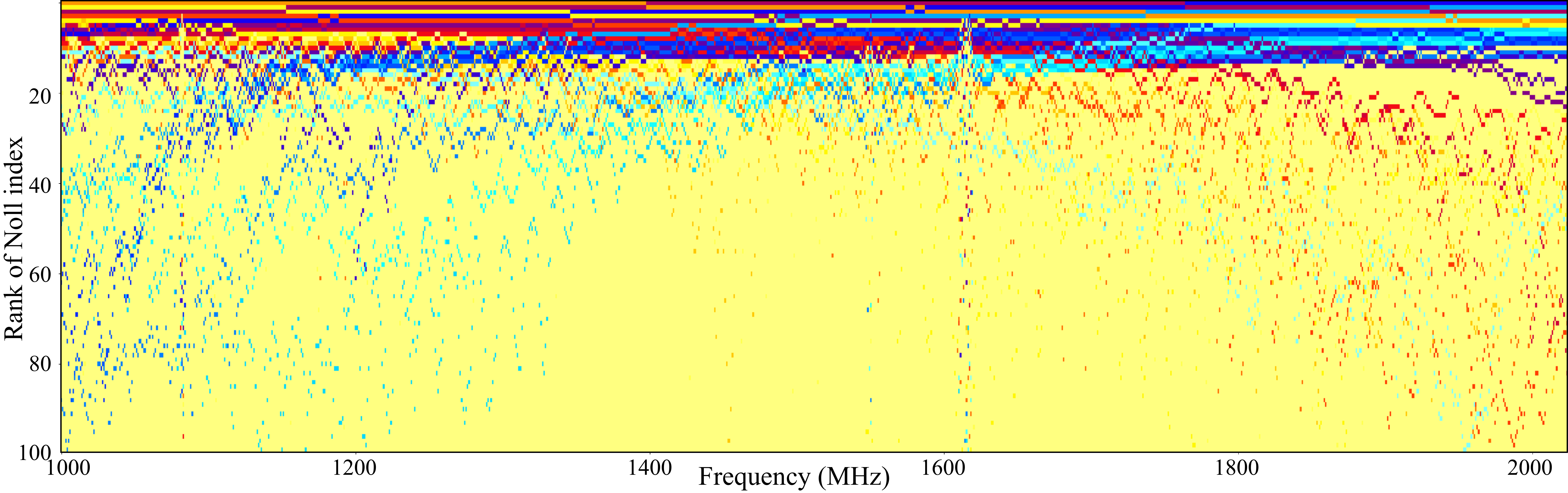}
\caption{Spectral behaviour of 23 Zernike dominant modes (Noll indices = 1,  2,  4,  7,  8, 11, 12, 14, 16, 17, 22, 26, 29, 30, 37, 40, 44, 47, 56, 60, 64, 79, 80) for antenna 6. The $y$-axis represent the rank from the most dominant (0$^{\text{th}}$) to the least dominant mode (100$^{\text{th}}$). Each trace hue is associated with a given Noll index. One can see the frequency domain where each specific Zernike mode is among the dominant modes that correctly represent the beam.}
\label{Fig11-Nollranknew}
\end{center}
\end{figure*}

For antenna 6, we plot in Fig. \ref{Fig10-ZERant6vectornew} the first 16 dominant Zernike polynomials that can encode the beam. Each Zernike mode is associated to its Noll index. A typical frequency varying beam can be represented with small (resp. large) Noll indices to encode the low (resp. high) frequency. At a given frequency, the beam exhibits a specific set of dominant Zernike modes. Due to the frequency scaling of the beam across the L-band, not the same set of Zernike will be useful to describe its shape.
%
%
If one wants to model the beam over the full band, it is required to monitor which Noll index is associated with the dominant modes.
To investigate this, we performed a Zernike decomposition of each channel over the first 100 Zernike modes, and we ranked the dominant Zernike mode in decreasing energy. The colour map of Fig. \ref{Fig11-Nollranknew} represents the distribution of the ranks of the first 23 dominant Zernike modes that were at least ranked among the first dominant 20. The y-axis is the rank order from 0$^{\text{th}}$ (= most dominant) to the 100$^{\text{th}}$ (= 100$^{\text{th}}$ most dominant or least dominant). Each of the tracked Zernike modes is represented by a single trace with a single colour. Small Noll indices are dominant at low frequencies and therefore occupy top ranks at the low end of the band. At high frequencies, those modes are no longer dominant and quickly lose ranks. Conversely, large Noll indices are dominant in the upper part of the band. Crossing over between large and small Noll indices occur around 1500 MHz.
%
%

To be able to represent the holography beams with enough accuracy, a selection of the relevant Noll indices is required. One could split the band into two bands and then have two sets of the most dominant Zernike modes in each sub-band.

\begin{figure}
\begin{center}
\includegraphics[width=\columnwidth]{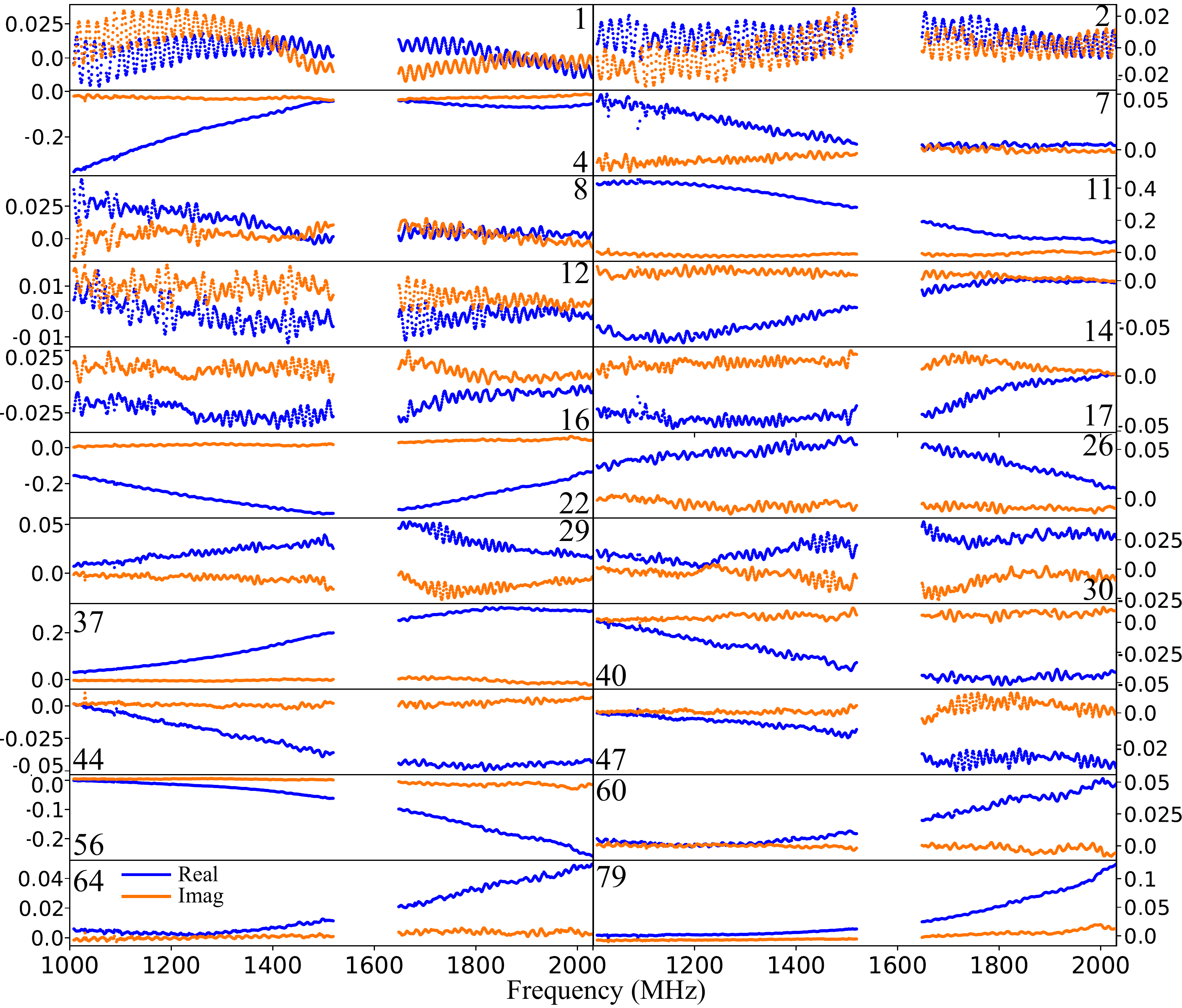}
\caption{Spectral behaviour of the coefficients associated to the first 22$^{\text{th}}$ dominant Zernike modes (real and imaginary) for antenna 6 that were ranked amongst the 20$^{\text{th}}$ across the L-band. The missing bands correspond to flagged data and were later reconstructed by the Papoulis-Gerchberg algorithm.}
\label{Fig12-ZERant6coeff22}
\end{center}
\end{figure}

Each of the selected Zernike modes has complex coefficients that have their own spectral behaviour. The coefficients are plotted in Fig. \ref{Fig12-ZERant6coeff22} and are numbered by their Noll index. To illustrate the previous discussion, the Noll indices 4, 11 and 14 modes are dominant at low frequencies whereas Noll indices 56, 64 and 79 are dominant at high frequencies. Some are dominant in the middle of the band (e. g. 22, 26).

It is interesting to note that the Noll = 1 mode is the piston term which corresponds to the DC component of the beam. Likewise, Noll = 2 encode an east-west tip-tilt error associated with pointing error. Noll = 3, encoding the north-south tip-tilt, is not among the dominant ones in the full band. This is consistent with $LL$ beam offset of antenna 6. At each frequency, the beam has a specific morphology, that is associated with a specific set of dominant Zernike modes. As the frequency increases, the beam smoothly scales down and the energy is transferred to subsequent Zernike modes.

\section{Spatial and spectral compression}

\subsection{Compressing the spectral information}
\label{compress1d}
The decomposition coefficients (given in a SVD/PCA or the Zernike orthonormal basis) of all holography beam cubes for all target antennas have been computed and stored for further experiment.
Regardless of the kind of spatial decomposition of the holography cube, we want to reduce the number of coefficients needed for adequate beam representation over the full L-band.
For each mode of each decomposition, we have a series of $N_\text{freqs}$ complex coefficients that describe the spectral distribution of this basis vector contribution. We now address the spectral encoding of these coefficients using 1D decomposition and low-rank approximations.

Radio Frequency Interference (RFI) is known to destroy the measured information in isolated (or groups of) channels, making them unavailable for data processing and data analysis. In our study of the spatial modelling of the holography cubes, we systematically discarded the 2D beams corresponding to a series of bad channels between frequencies 1519 MHz and 1648 MHz in order to avoid their non-physical contribution to the different decompositions. Contrary to the SVD/PCA which operates on all good channels, the Zernike decompositions were performed on a channel-by-channel fashion and are therefore less affected by bad channels. The bad channels were discarded anyway, leaving a large gap in the coefficients.

\subsubsection{Restoring missing channels}
\begin{figure}
\begin{center}
\includegraphics[width=\columnwidth]{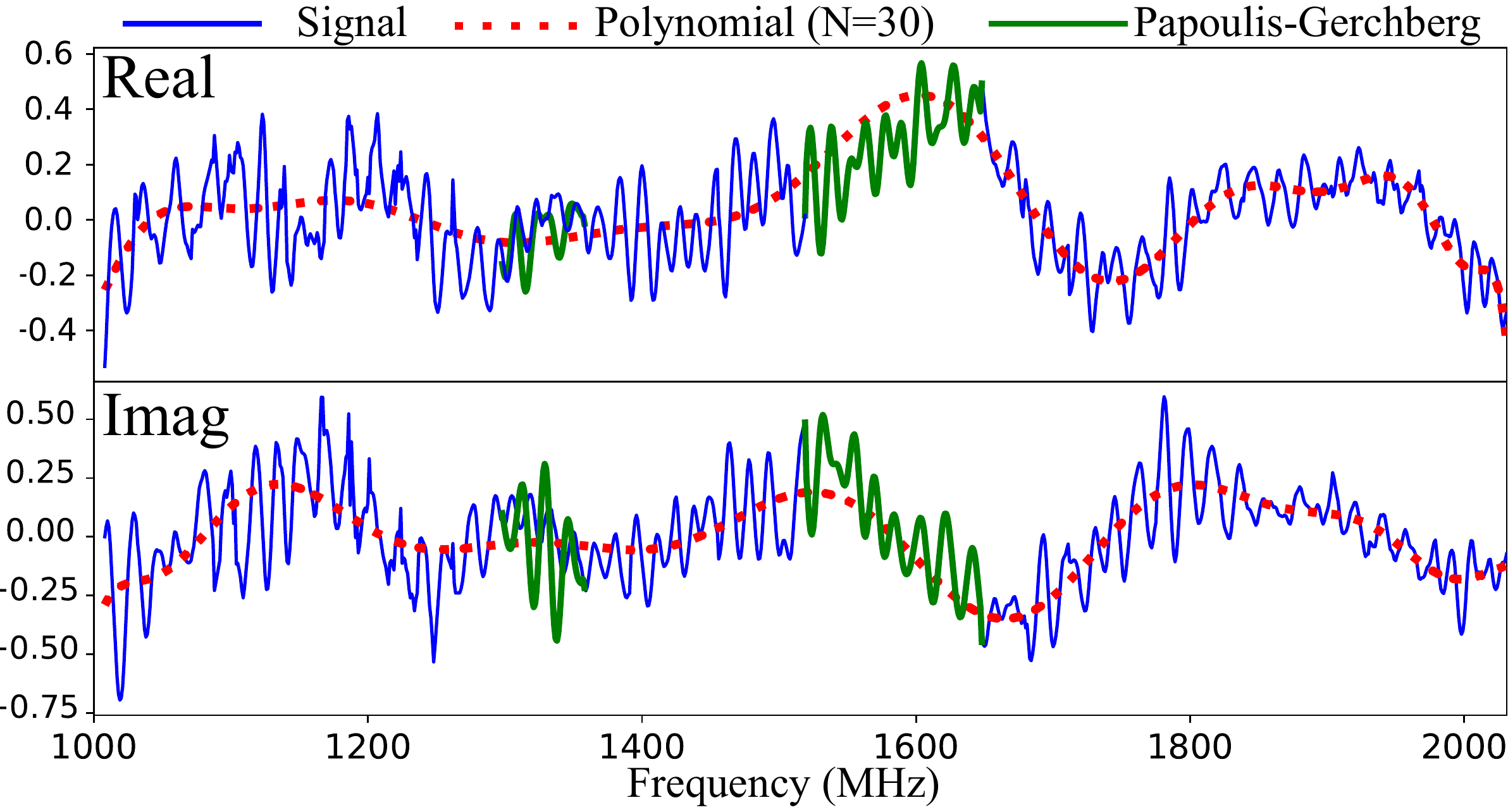}
\caption{Signal restoration with the Papoulis-Gerchberg algorithm used with DCT decomposition in the 1519--1648 MHz range: the original signal, its polynomial interpolation of order 30 and Papoulis-Gerchberg restoration (bold). The 1300--1350 MHz range was also restored and compared to the signal.}
\label{Fig13-Papoulis}
\end{center}
\end{figure}

Figure \ref{Fig13-Papoulis} displays the real part and imaginary part of the coefficients associated with the first eigenvector of the SVD/PCA decomposition of antenna 6 holography cube. We can see a fast oscillating trend corresponding to the frequency ripple superimposed onto a smooth low-order trend. In order to encode this information in frequency, we will rely on the Discrete Cosine Transform (DCT).
For all coefficients, however, the lack of information corresponding to the bad channels prevent us from obtaining information on the beam in these channels. To first order, we can compensate for this by reconstructing the missing channels with linear or polynomial interpolation. However, the interpolated regions will not account for the ripple which is known to be present over the full L-band. If we decompose this signal with a DCT (or Fourier transform, or a wavelets transform), the corresponding decompositions will suffer from discontinuities at the interpolated region borders. These radical changes in shape (especially if it implies discontinuities) will cause a Gibbs-like phenomenon in the interpolated regions under low-rank approximations (e.g. see the Fourier approximation of a step function and its successive low-rank approximations). When no interpolation is used, a low-rank DCT tries to represent a plateau of zero-valued coefficients (corresponding to the missing segment). With linear and polynomial interpolation, a low-rank DCT reproduces the slow variations but strongly biases the amplitude of the ripple in the missing segment interval.
A better way to interpolate the missing information that enables a robust DCT decomposition is to rely on signal restoration methods. We implemented the well-known and fast Papoulis-Gerchberg algorithm (PGA  -- \cite{Papoulis_1975}) which is an iterative thresholding algorithm which performs signal restoration by \textit{inpainting}. It is designed to interpolate band-limited signals. We make the assumption that the frequencies contained in our signal are band-limited (the highest-frequency being a bit larger than that associated with the ripple). We have used the PGA in conjunction with the DCT transform itself. Figure \ref{Fig13-Papoulis} illustrates the successful restoration of the missing information compared to linear and polynomial interpolations. On one hand, linear interpolation highly depends on the last and first good coefficient before and after the gap. On the other hand, a low-degree polynomial interpolation could reconstruct the trend but not the rapid oscillations of the ripple. We applied the PGA to each decomposition coefficients and we used a maximum of 3000 iterations to converge, keeping the 50 most powerful DCT modes.

In order to test the robustness of the method, we blanked known the 1300--1350 MHz band for which we know the ground truth. Perfect reconstruction would make coincide the signal curve with the PGA reconstruction curve. However, even if the fast variations is well approximated, the low level variations is not optimal. This shows that the default PGA is sufficient to help replacing the missing channels by approximated estimates of the ground truth, but it cannot serves as a method to derive `exactly' the missing underlying signal. As presented earlier, we aim at compressing 1D signal with DCT coefficients with the smallest possible bias. However, having missing channels is equivalent to applying a sharp mask over the full signal, leading to strong bias in the transformed space over many DCT coefficients (and therefore, more costly to store). In order to lower this bias, we used PGA to get an approximation of the missing data along with selecting the most representative DCT coefficients of the signal. If one aims at low-bias compression, the PGA reconstruction is sufficient. More robust reconstruction can be done by enforcing the sparsity on the reconstruction (see \cite{Kayvanrad_2009}) but it was not applied in our reconstruction.

The final `interpolated' signal reproduces the main trend of the signal as well as the fast varying features. We now have a method that enable the selection of the most representative features of the signal. In our study, PGA was applied on SVD/PCA as well as on the Zernike coefficients.
%
%

\subsubsection{1D Compression for all representations}

After signal restoration, we want to further compress the spectral information of all coefficients. The final representation of the beam will, therefore, have two stages: i) spatial encoding and approximation of the beam on a basis of representative basis functions and ii) the spectral encoding of the corresponding coefficients along frequency.
The spatial encoding is either done with SVD/PCA or Zernike decomposition. The spectral encoding was performed using a DCT. We selected a maximum number of 50 DCT modes.
To evaluate the `compressibility', we counted the total number of coefficients that are required to represent the full beam at any size.
The full-Stokes holography beam cube can be represented by a complex-valued matrix of size $4 \times N_{\mathrm{freq}} \times N_{\mathrm{pix}}^2$.
Let us assume that we want a $(k_{lm},k_\nu)$ rank approximation of the beam, where $k_{lm}$ and $k_\nu$ are respectively the cut-off rank for spatial and spectral information modelling.
SVD requires the additional knowledge of the matrix $v_{k_{lm}}^T$ (of dimension $k_{lm}\times N_{\mathrm{pix}}\times  N_{\mathrm{pix}}$) which represents the first $k$ dominant eigenbeams. The corresponding $u_{k_{lm}} w_{k_{lm}}$ (of dimension $k_{lm}\times N_{\mathrm{freq}}$) are the coefficients as a function of frequency. If each coefficient can be modelled with $k_\nu$ DCT modes, the total amount of coefficients to represent one antenna polarization can be depicted as follows:
$$N_{\mathrm{SVD}} =2\times\underbrace{k_{lm}\times k_\nu}_{\text{coeffs}}+\underbrace{k_{lm}\times N_{\mathrm{pix}}\times  N_{\mathrm{pix}}}_{\text{eigenbasis } v_{k_{lm}}^T }$$

$$N_{\mathrm{Zernike}} =2\times\underbrace{k_{lm}\times k_\nu}_{\text{coeffs}}+\underbrace{k_{lm}}_{\text{list of Noll indices}}$$
where the factor 2 is to count real and imaginary parts of the coefficients. 

In addition to the $k_{lm}\times k_\nu$ coefficients, the Zernike decomposition requires the list of the $k_{lm}$ Noll indices associated with the $k_{lm}$ coefficients.

The compression factor $CF$ can be computed as the ratio of $(N_{\mathrm{freq}}\times N_{\mathrm{pix}}^2)/N_{\text{2D}}$ where $N_{\text{2D}}$ is the number of spatial scales (i.e. $N_\text{SVD}$ or $N_\text{Zernike}$).

For J2018, one can actually compress these coefficients in frequency while maintaining the other parameters fixed:
$$N_{\mathrm{Cassbeam}}=  \underbrace{7\times k_{\nu}}_{\text{Fig. \ref{Fig7-Preshantparam}}} + \underbrace{N_{CB}}_{\text{other fixed model params}}$$

When using the HT step on the inverse FT of the holography beam, we can store the pixel coefficients corresponding to the aperture plane. Due to the frequency dependency of the FFT, one has to store respectively 30, 37 and 57 complex coefficients at 1008, 1408 and 1908 MHz. By approximating the increase of the number of coefficients as a linear dependency with frequency, we need $\approx$1040 coefficients to approximate the data in all 1024 channels. Considering the low quality of reconstruction, this total number of coefficients comes with a lower value of compressibility, thus HT denoising is not a prefered solution to compress the holography beam information.

\subsection{Reconstruction accuracy vs. compressibility}
\begin{figure}
\begin{center}
\includegraphics[width=\columnwidth]{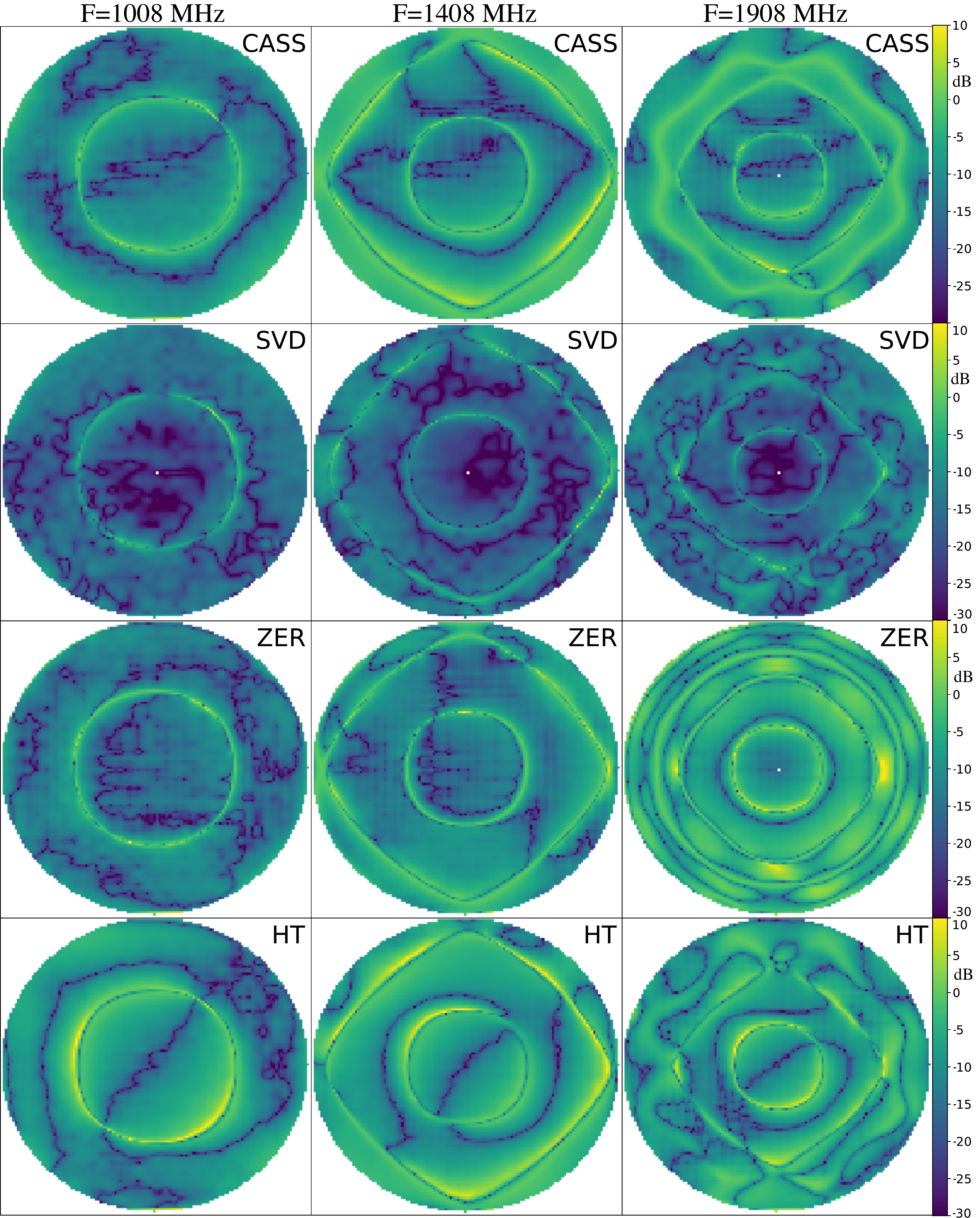}
\caption{Relative residual error represented with the same colour bar coding the range [-30 dB, 10 dB] at F=1008, 1408, 1908 MHz from the difference of the original holography beam and CASS ($RR$) J2018, SVD ($RR$), ZER ($LL$) and HT ($LL$) reconstructed beams. Reconstruction quality with $RR$ and $LL$ beams are equivalent.}
\label{Fig14-comparisonresiduals}
\end{center}
\end{figure}

We compared all these representations against one another in terms of accuracy and number of coefficients. All of them have advantages and drawbacks depending on the application. The series of `compressed' coefficients can be decompressed again to restore the whole series of $N_{\mathrm{freq}} \times k_{lm}$ coefficients. Then, the approximated beam cubes can be constructed back with the linear sum of the coefficients with the basis vector.
To compare the quality of all representations, we compared the relative reconstruction accuracy using the Normalized Root Mean Square Error (NRMSE) as the figure of merit.
We could not directly reproduce the data from J2018, but we obtained a set of optimized Cassbeam beams at regularly-spaced frequencies from the authors. We present in Fig. \ref{Fig14-comparisonresiduals}, single-channel comparison of residuals for antenna 6 of the SVD/PCA and Zernike approximations using $k_{\nu}=50$ and $k_{lm}=20$ with the original holography beams and the fitted Cassbeam provided by J2018. This figure depicts the $RR$ and $LL$ beam reconstruction residuals (relative to the original holography measurements) at frequencies 1008 MHz, 1408 MHz and 1908 MHz for the four representation addressed in this paper: Fitted Cassbeam from J2018 (`CASS'), the SVD decomposition (`SVD'), the Zernike decomposition (`ZER') and the HT denoising (`HT'). 
The low-frequency beam is correctly modelled by all four representations. Fitted Cassbeam and Zernike decomposition show similar performance with around -10dB of average error. HT is the worst representation, which also comes with the largest number of coefficients.
Nulls and sidelobes are generally poorly represented in all decompositions and are associated with large errors. This effect is mostly due to the truncation of the coefficients in the representation which tends to misrepresent the region of the nulls.
%
%

From Fig. \ref{Fig14-comparisonresiduals}, one can notice that SVD is the most robust way to represent the beam over all the bands (average of -15 dB). This is expected as the eigenbases are directly derived from the data.


\begin{figure}
\begin{center}
\includegraphics[width=\columnwidth]{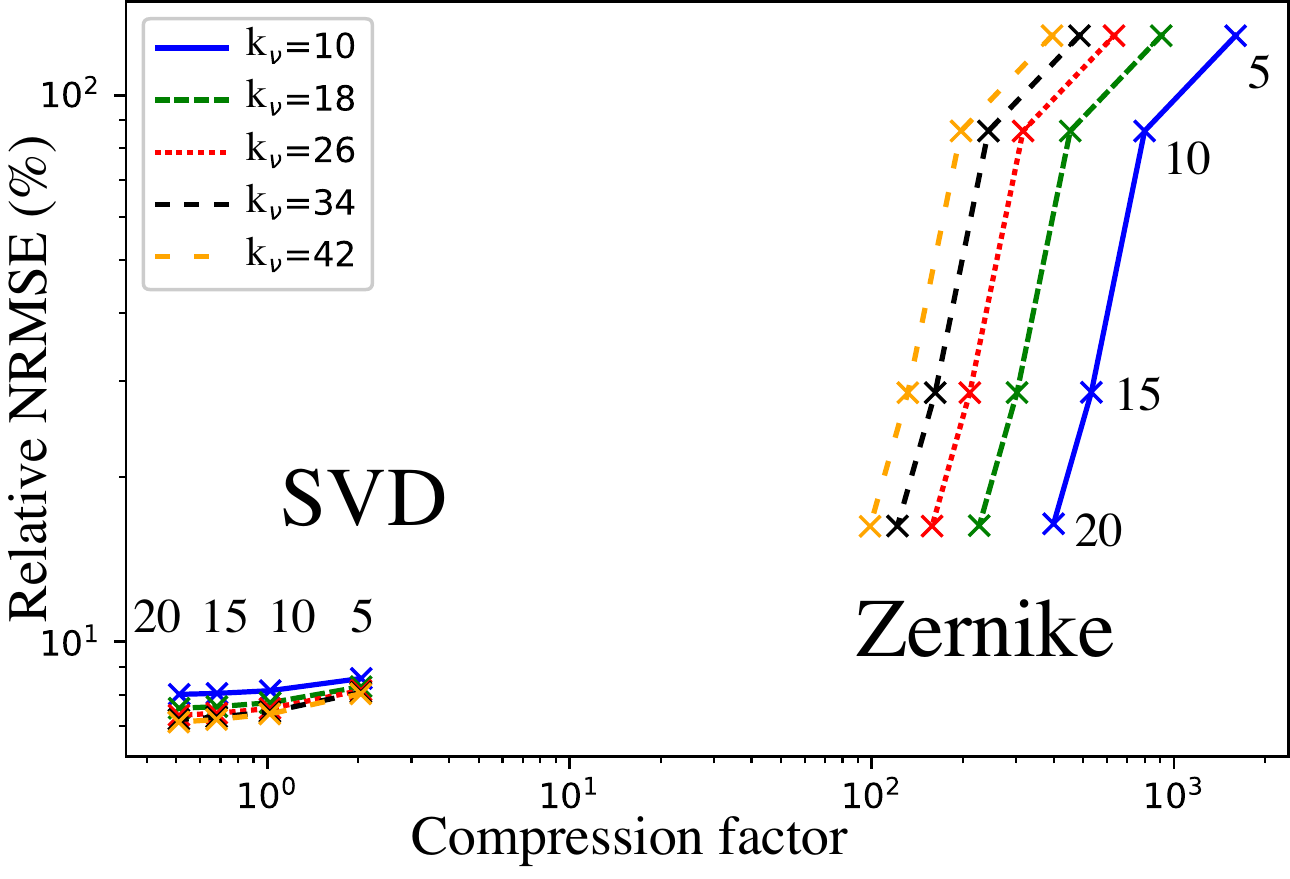}
\caption{NRMS error of SVD/PCA and Zernike reconstructions compared to the original holography cube for $k_{\nu}=[10...50]$ and $k_{lm}=[5...20]$}
\label{NRMSE}
\end{center}
\end{figure}

In order to measure the overall reconstruction error vs. compressibility, we reconstructed the holography beams for $k_{\nu}=[10...50]$ and $k_{lm}=[5...20]$. We plot the NRMSE as a function of the compression factor with the original holography cube in Fig. \ref{NRMSE}. This error represents the relative variation w.r.t. the holography cube, but this plot has to be taken with precaution.
For both decompositions, one can note that the variations of $k_\nu$ from 42 to 10 do not impact the relative NRMSE but do impact the compression factor by almost one order of magnitude. However, variations of $k_{lm}$ have a different impact on SVD and Zernike reconstruction error.

SVD decomposition offers the best NRMSE values but comes with lower compression factor values due to the necessity to store the $k_{lm}$ first eigenvectors (as $N_{\mathrm{pix}}^2$ matrices), however, SVD decomposition can lead to overfitting the features associated with the holography measurement procedure itself (those features can be spotted as vertical and horizontal tracks in the CASS and ZER residuals of Fig. \ref{Fig14-comparisonresiduals}.

The Zernike decomposition offers larger compression factors than SVD, but less accuracy over the full band. This is due to the difficulty of selecting the relevant Zernike modes that represent the beam over the full band. a small number of modes translates into severe NRMSE error.
However, channel-to-channel Zernike decomposition displays percent-level errors on average.

Similar trends can be noticed for both $RR$ \& $LL$ but lower quality fits and compressibility results are noted for $RL$ \& $LR$ beams. Decomposition coefficients and bases can be found on \url{https://github.com/kmbasad/eidos}.


\section{Conclusions}
\label{sec-conclusion}

Advanced calibration methods that include DDE are vital to enhancing the quality of interferometric images (taking most of the instrument sensitivity), improving the survey speed and enabling HDR imaging beyond the first beam null.
To enable that, accurate beam modelling is required and EM models have shown their limits as they do not model `on-site' variations that are not included in the theoretical models.

We present in this paper, an approach to model the VLA antenna in the L-band using holography data. We searched for adapted representations that can accurately and efficiently represent the beam and its spatial and spectral variations.

We used the data presented in J2018 but generalizes it to all 12 measured antennas.

In order to capture the mean features of the beam and its dependence in frequency (pointing error, FWHM scaling and ripples), we used dedicated methods to model the beam in 3D (space and frequency).

J2018 proposed an updated version of the Cassbeam ray-tracing beam model which consists of fitting physical parameters of the antenna to the holography measurements. Apart from the well-known pointing offset and squint, a noticeable 17.2 MHz ripple can be observed across the band as in J2018. The authors suggest that this ripple effect can be modelled by the variation of the central aperture blockage parameters. However, we have shown that other physical parameters of this model are also strongly dependent on the ripple frequency, limiting the explanatory power of interpreting the ripple in a limited number of physical parameters.

Our study was motivated by `model-independent' representation of the beam. We confronted two characteristic approaches to model the beam: i) a `data-driven' basis decomposition (with SVD/PCA) and ii) a projection of the data on the Zernike orthogonal basis. This decomposition accounts for the spatial variations of the beam at each frequency. We derived the frequency distribution of decomposition coefficients of the most dominant modes in various basis. To model the frequency behaviour of each decomposition modes, we compressed the information using DCT after the reconstruction of the missing channel information with the PGA.

We have seen that the `data-driven' and `orthogonal projection' come with advantages and drawbacks due to their nature: 
\begin{itemize}
\item SVD/PCA provide accurate reconstructions even with low-rank approximations, due to the fact that the associated basis is the best representing the data. However, SVD/PCA is `rigid', in the sense that it requires storing the low-rank set of eigenvectors $v_{k_{lm}}^T$ of the decomposition for a given support size $N_{\mathrm{pix}}^2$ in addition to the coefficients. The SVD/PCA procedure does not require a channel-by-channel analysis as it considers the whole 3D holography cubes as a whole. It can be a problem if the data are not properly flagged.

\item The Zernike decomposition provides an adequate set of vectors to represent the main features of the beam in a generic way. In addition, we only need to have a compressed set of coefficients with Zernike basis and we do not need to compute the basis with SVD/PCA. Therefore, the beam can be regenerated at any arbitrary size depending on the application. The channel-by-channel Zernike decomposition makes this method less subject to RFI compared to SVD/PCA for which the bad channels need to be taken out before computing the eigenbasis and the corresponding coefficients.
One drawback of the Zernike decomposition is the difficulty in selection of the dominant modes that can represent the beam over the full L-band.

\item In order to have an even sparser representation of the spatial and spectral behaviour of the beam, we compressed the spectral information using DCT low-rank approximations. DCT was preferred to other transforms due to the fact that slow and smooth variation and the frequency ripple over the whole L-band can efficiently be modelled in much fewer coefficients that the number of channels. This second-stage DCT decomposition in frequency encodes the slow and fast variations and substantially improves the compression factor of the model.
The final model can be represented by a very low number of coefficients (depending on the target fidelity level).

\item The coefficients of the antenna model parameters derived by J2018 can also be compressed in frequency with DCT. However, to compare the modelling performance of our methods with that of J2018, we directly compared the beam reconstruction residuals with that obtained from J2018 work.

\item With the currently available data, we could not study the dependence of the primary beam on observation parameters such as the elevation angle (due to the gravitational load of the antenna structure). This requires additional systematic holographic observations taken when the source is at lower elevations. For the MeerKAT antenna (addressed in the next paper), different elevations were observed.

\end{itemize}
Based on the comparisons of the beam reconstruction residuals, we could conclude that :
i) SVD/PCA is better than any other method over the L-band (including J2018), ii) our decomposition methods do not require any a priori knowledge of the geometry of the antenna, iii) we do not require any specific physical or mechanical interpretation of the ripple to model it (even J2018 explanation is not related to the real origin of the ripple.). Zernike polynomials can be interpreted in terms of optical distortions (e.g. offset, tip/tilt, astigmatism, coma, etc.) even if it would be more appropriate to study the aperture plane (as shown in the next paper of the series).

Future work will address beam modelling in all VLA bands based on the same methods. In the scope of a long-term monitoring of the VLA, annual holography campaigns in all frequency bands can be performed in order to monitor each antenna and model the effective associated response to guarantee their maximum sensitivity.
Having an accurate beam model that includes refined instrumental effects is of primary importance when performing large-scale surveys (such as the VLA Sky Survey), in particular, in the S-band. The ripple effect seen at L-band is present also in S-band, despite at a smaller amplitude.  Indeed, large FOV imaging is possible even beyond the first beam null. This can dramatically reduce the survey speed at the condition of performing direction-dependent calibration with an accurate beam model. The advanced calibration improves the quality of the calibration solutions and lead to higher dynamic range images and better accuracy of the observed flux density in catalogues. Better calibration also reduces systematic calibration artefacts in images, enabling an interferometer to reach is theoretical sensitivity by discovering faint sources.
The sequel of this study will be address in paper II with the application of our methods to the new offset-Cassegrain antenna currently in use in MeerKAT. Next, paper III especially addresses the impact of the quality of the beam model (and of the ripple) to that of the calibration and imaging in the context of HI imaging. These topics remains of primary importance in the context of modern radio interferometric imaging.

\section*{Acknowledgements}
The research of KI and OS are supported by the South African Research Chairs Initiative of the Department of Science and Technology and National Research Foundation. JG acknowledges the financial support from the UnivEarthS Labex program of Sorbonne Paris Cit\'e (ANR-10-LABX-0023 and ANR-11-IDEX-0005-02) and from the European Research Council grant SparseAstro (ERC-228261).


\bibliographystyle{mnras} 
\bibliography{refs} 

\end{document}